\documentclass[reprint,
 amsmath,amssymb,
 aps,
]{revtex4-2}

\usepackage{soul}
\usepackage{esint}
\usepackage{dcolumn}
\usepackage{bm}

\usepackage[T1]{fontenc}
\usepackage{amsmath,amsfonts,amssymb}
\usepackage{nccmath}
\usepackage{physics}
\usepackage{braket}
\usepackage[english]{babel}
\usepackage{graphicx}
\usepackage{here}
\usepackage[urlcolor=red,linkcolor=blue,colorlinks=true]{hyperref}
\usepackage{float}
\usepackage{url}

\begin{document}
\title{Effects of atom losses on a one-dimensional lattice gas of hardcore bosons}

\author{Fran\c {c}ois Riggio, Lorenzo Rosso$^{\dagger}$, Dragi Karevski and J\'er\^ome Dubail}
\affiliation{{}$^{\dagger}$Universit\'e Paris-Saclay, CNRS, LPTMS, 91405 Orsay, France \\ Universit\'e de Lorraine, CNRS, LPCT, F-54000 Nancy, France }
\email{francois.riggio@univ-lorraine.fr}

\begin{abstract}
Atom losses occur naturally during cold atoms experiments. Since this phenomenon is unavoidable, it is important to understand its effect on the remaining atoms. Here we {\color{black} investigate a toy model: a lattice gas} of hard-core bosons subject to $K$-body losses (where $K=1,2,3,\dots$ is the number of atoms lost in each loss event) on $K$ neighboring sites. In particular we investigate the effect of losses on the rapidity distribution $\rho(k)$ of the atoms. Under the assumption that losses are weak enough so that the system relaxes between two loss events, we are able to determine the loss functional $F[\rho](k)$ encoding the loss process for $K$-body losses. We derive closed expressions for the cases of one- and two-body losses, and show their effects on the evolution of the total number of particles. Then we add a harmonic trapping potential and study the evolution of the position-dependent rapidity distribution of this system by solving numerically the evolution equation for one-, two- and three-body losses.
\end{abstract}

\maketitle
\section{Introduction}
During the past two decades, cold atom experiments have become a key simulation platform to investigate the physics of low-dimensional quantum many-body systems~\cite{Bloch2008}. While these experiments are typically well isolated and their dynamics at short times is well approximated by unitary dynamics, cold atom setups are always subject to atom losses. The effects of the latter on the dynamics of the gas can become important at longer times, and they are typically hard to describe theoretially. Cold atom experiments involve various loss processes, for instance one-body losses resulting from scattering with background thermal atoms~\cite{Knoop2012}, which can be significant. Inelastic two-body collisions can occur naturally or be intentionally engineered, leading to two-body losses~\cite{Browaeys2000,Franchi2017,Tomita2017,syassen, Tomita2019,Yan2013,sponselee, Honda2023}. Three-body losses, where a highly bound diatomic molecule is formed, are invariably present and typically dominate the overall loss process~\cite{Soding1999, Weber2003, Bruno2004}. In principle, loss processes involving more than three atoms also exist. In particular, losses involving four atoms have been reported in Refs.~\cite{Ferlaino2009,Gurian2012}. Loss processes can sometimes be controlled and engineered~ \cite{harber_thermally_2003} to bring out physical phenomena such as cooling~\cite{johnson_long-lived_2017,Bouchoule2018_4, Dogra_2019,Grisins_2016,Rauer2016,Max2017,Schemmer2018,Schemmer2020_4} and the quantum Zeno effect~\cite{rossini_strong_2021,garcia-ripoll_dissipation-induced_2009,Misra1977,Itano_1990,Beige_2000,Almut_2000,Kempe_2001,Facchi_2001,Facchi_2002,Schuetzhold_2010,
Stannigel_2014,Gong_2017,Froml_2019,Snizhko_2020,Biella_2021}.

Despite the relevance of loss processes, a consistent general theory describing the effects of losses on the dynamics does not exist yet.
One standard way to model atom losses is to use the Lindblad equation \cite{garcia-ripoll_dissipation-induced_2009,bouchoule_effect_2020,rossini_strong_2021,bouchoule_losses_2021, minganti2023dissipative}. Several studies have inspected the interplay between the unitary dynamics and the lossy one in both bosonic~\cite{garcia-ripoll_dissipation-induced_2009, rossini_strong_2021, bouchoule_effect_2020, rosso_one-dimensional_2022, huang2023modeling} and fermionic gases~\cite{Zhu2014,rosso_one-dimensional_2021-1, Rosso4,Rosso5,Nakagawa2020,Nakagawa2021, Perfetto2023, huang2023modeling}.

{\color{black} In one spatial dimension, several ultracold gases are integrable or nearly integrable, including continuous gases of bosons~\cite{paredes2004tonks,kinoshita2004observation,haller2009realization,van2008yang,jacqmin2011sub} or fermions~\cite{moritz2005confinement,liao2010spin}, sometimes with multiple components~\cite{pagano2014one,wicke2010controlling}, but also lattice gases like hard-core bosons~\cite{ronzheimer2013expansion,vidmar2015dynamical} or the Fermi-Hubbard model~\cite{essler2005one}. The question of losses in those integrable gases is particularly interesting.} An integrable model admits a macroscopic number of conserved charges in comparison with non-integrable models where, typically, only the energy and the number of particle are conserved. Due to this large set of conserved quantities, an isolated integrable system has a singular property: the stationary states of the system are modeled by Generalized Gibbs Ensembles (GGE). A GGE is constructed by maximizing the entropy under the constraints imposed by all conservations laws \cite{doyon_lecture_2020}. However, under dissipative processes like atom losses, the conserved quantities of an integrable system are not conserved anymore, as the coupling between the system and its environnement typically causes integrability breaking \cite{hutsalyuk_integrability_2021}. In principle, one then expects that integrability breaking leads to thermalization at very long times, possibly with a pre-thermalization phenomenon at intermediate time scales~\cite{bertini_prethermalization_2015}. Some studies seem to confirm the thermalization  but other works suggest otherwise \cite{johnson_long-lived_2017}.

\begin{figure}[tb]
    \centering
    \includegraphics[scale=0.45]{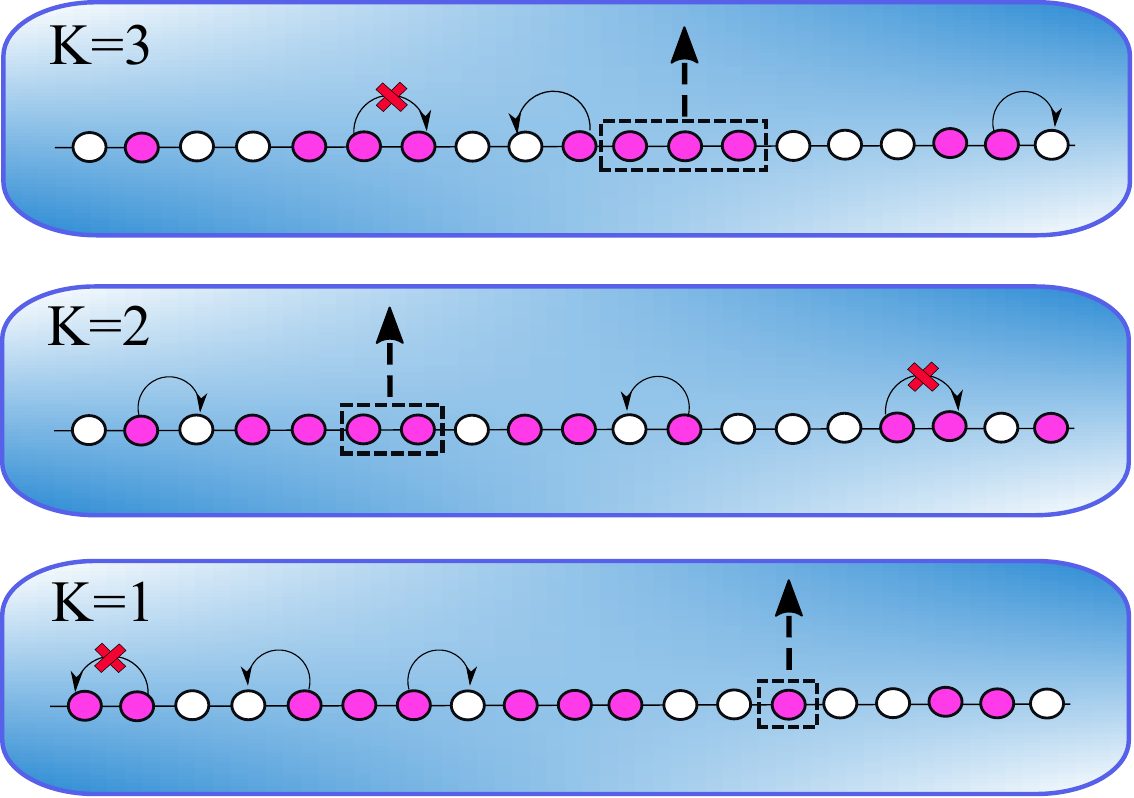}
    \caption{We consider a lattice gas of hardcore bosons (pink dots) that can jump from site to site. Two bosons cannot be on the same site. Losses are represented by the large black arrow which connects the bosons chain to its environment (blue background). Here we show the situation for one-body losses ($K=1$), two-body losses ($K=2$) and three-body losses ($K=3$). For each cases, $K$ consecutive atoms are removed and are lost in the `environment'. In this paper losses are modeled by the Lindblad equation (\ref{LindbladBH}).}
    \label{schema_loss}
\end{figure}

{\color{black} Motivated by the outstanding challenge of developping a general theory of the effects of losses on the dynamics of integrable gases, here we introduce a toy model where  we can theoretically investigate the effects of $K$-body losses ($K \geq 1$) on the dynamics of the simplest such gas. We propose to investigate hard-core bosons on a lattice ---a famous model of integrable quantum gas dynamics that has been extensively studied theoretically~\cite{rigol2005fermionization,rigol2006hard,rigol2007relaxation,vidmar2013sudden,vidmar2017emergent,xu2017expansion} and realized experimentally~\cite{ronzheimer2013expansion,vidmar2015dynamical}--- subject to weak $K$-body atom losses: whenever $K$ atoms occupy $K$ neighboring sites, they can escape the system with some rate (Fig.~\ref{schema_loss}).

We stress that experimentally realistic loss processes in optical lattices usually involve atoms on the same lattice site, so, strictly speaking, only the $K = 1$ version of our toy model is of experimental relevance. For $K\geq 2$, we are not aware of existing experimental setups that could engineer our $K$-site loss processes. However, from a theoretical perspective this model is very natural.} Here our goal is to understand how integrability breaking caused by such loss events affects the dynamics of the gas. Of course, under atom losses, the stationary state at very long times is always the vacuum. However, what we want to investigate is how the vacuum is approached, and in particular whether or not to the gas is in a quasi-stationary thermal state. We investigate for instance the mean density $n(t)=\braket{N(t)}/L$ (where $N$ is the number of atoms and $L$ is the number of lattice sites) for different initial states and observe different non-trivial behaviors depending on the number $K$ of atoms lost in each loss event. In particular, for two-body losses ($K=2$), we analytically show that $n(t)\propto 1/t$ or $\propto 1/t^{1/2}$, depending on a specific property of the initial state (see below). More generally, for other values of $K \geq 2$ our numerical results are in agreement with a power-law decay $n(t) \propto t^\alpha$ where the exponent $\alpha$ typically depends on the initial state, and generically differs from the mean-field result $1/(K-1)$.  This observation holds also if we add a harmonic potential; however in that case the exponent $\alpha$ changes and is different from the one found in the homogeneous case.

The paper is organized as follows. In Section~\ref{sec:model} we define the model and discuss our assumptions. Our main hypothesis of slow losses, and fundamental concepts such as the rapidity distribution and the loss functional are introduced. In Section~\ref{sec:lossfunctional} we present our calculation of the loss functional for $K$-body losses, which crucially depends on the parity of $K$. In Section~\ref{sec:homogeneous} we investigate the effect of atom losses on the rapidity distribution and on the mean density in the homogeneous gas, by combining analytical and numerical calculations. In Section~\ref{sec:trapped}, we add a harmonic trapping potential. We use a hydrodynamic-like approach similar to Generalized Hydrodynamics~\cite{castro-alvaredo_emergent_2016,bertini_transport_2016} where we incorporate the loss functional (following similar proposals in Refs.~\cite{bouchoule_effect_2020,rosso_one-dimensional_2022}), and design a numerical method for solving the resulting evolution equation for the rapidity distribution in the gas. We conclude in Section~\ref{sec:conclusion}.

\section{The model}
\label{sec:model}
\subsection{Definition: lattice Tonks-Girardeau gas with $K$-body losses}

We consider a lattice Tonks-Girardeau gas subject to atom losses. 
Each site $j \in \mathbb{Z}$ is occupied by either zero or one boson. We write $\sigma^+_j$/$\sigma^-_j$ for the operator that creates/annihilates a boson on site $j$. Because of the hard-core constraint, these operators do not satisfy the usual bosonic canonical commutation relations. Instead, they satisfy the algebra of Pauli matrices, 
$(\sigma^+_j)^2 = (\sigma^-_j)^2 = 0$, and $ [ \sigma^+_i, \sigma^-_j ] = \delta_{i,j} \sigma^z_j$.

We consider the hard-core boson (HCB) Hamiltonian with nearest-neighbor hopping,
\begin{align}\label{HCB}
H_{\rm HCB}=- \frac{1}{2} \sum_{j \in \mathbb{Z} }(\sigma^{+}_{j} \sigma^-_{j+1}+ \sigma^{+}_{j+1} \sigma^-_{j}). 
\end{align}
This Hamiltonian generates the unitary part of the evolution of the gas. In addition, we assume that the gas is subject to incoherent $K$-body loss processes, with $K$ a positive integer. To describe the losses, we assume that the dynamics is Markovian, and we consider the following Lindblad equation for the density matrix $\hat{\rho}$,
\begin{eqnarray} \label{LindbladBH}
\nonumber \dot{\hat{\rho}}(t)&=& -i [H_{\rm HCB}, \hat{\rho}(t)] \\
 && +\Gamma \sum_{j \in \mathbb{Z}} \left( L_{j} \, \hat{\rho}(t) \, L^{\dagger }_{j}-\dfrac{1}{2} \{L^{\dagger }_{j} L_{j}, \hat{\rho}(t)\} \right),
\end{eqnarray} 
Here $\Gamma$ is a constant that sets the loss rate, while the Lindblad operators
\begin{align}\label{Lj}
L_{j}=\prod^{K-1}_{l=0} \sigma^-_{j+l}
\end{align}
remove $K$ bosons from the $K$ consecutive sites $j, j+1, \dots, j+K-1$.

We stress that, with these loss terms, the model is not exactly solvable, so it is necessary to develop some effective approaches to tackle it.

\subsection{Adiabatic losses, effective description by slow motion of the charges}

To simplify the description of the system, we follow the approach of Ref.~\cite{bouchoule_effect_2020}. There the approach was developed for losses in a continuous gas, and it is based on the assumption that the losses are slow, so that the gas remains in a Generalized Gibbs Ensemble (GGE) with parameters that slowly drift in time (see e.g.~\cite{bouchoule_generalized_2021} for a review). A similar description of slowly-evolving nearly integrable systems with weak integrability breaking had previously appeared in Refs.~\cite{lange_time-dependent_2018,Lange_2017,Lenarcic_2018}. Following these ideas, here we also assume that the loss processes occur on very long times compared to the relaxation time scale due to the unitary evolution of the gas. In that limit, the gas has time to reach a local stationary state after each loss event. To efficiently exploit that assumption, we look at the slow dynamics of the conserved charges.

For convenience, from now on we focus on a finite system of $L\gg 1$ sites, with periodic boundary conditions. The Hamiltonian $H_{\rm HCB}$ commutes with an infinite set of hermitian operators $Q_a$, $a = 0,1, 2\dots $ that can be constructed using the Jordan-Wigner mapping to non-interacting fermions (see Subsec.~\ref{subsec:rapidity} for details). These operators also commute among themselves, $ [Q_a, Q_b] = [H_{\rm HCB}, Q_a] = 0$. Moreover, they are local in the sense that $Q_a = \sum_{j =1}^{L} q_{a,j}$ where $q_{a,j}$ is a charge density operator that has compact support (i.e. it acts on a finite number of sites around $j$).

The time evolution of the expectation value $\left< Q_a \right>(t) = {\rm tr} [ \hat{\rho}(t) Q_a]$, is obtained from Eq.~(\ref{LindbladBH}),
\begin{align}\label{evoQ}
\dot{\expval{Q_a}}(t)= \frac{\Gamma}{2} \sum_{j =1}^{L} \expval{   L^\dagger_{j} \, [Q_a, \, L_{j}] + [ L^\dagger_{j} ,  \, Q_a ] L_{j} }(t).
\end{align} 
Moreover, the hermiticity of $Q_a$ implies $\langle  [L_{j}^\dagger, Q_a]  L_{j} \rangle = \langle   L^\dagger_{j} \, [Q_a, \, L_{j}] \rangle^*$. Thus,
\begin{align}\label{evoQ2}
\dot{\expval{Q_a}}(t)= \Gamma \sum_{j=1}^{L} {\rm Re} \, \expval{    L^\dagger_{j} [  Q_a , L_{j} ] }(t).
\end{align} 
This equation is exact and it is a direct consequence of Eq.~(\ref{LindbladBH}). Notice that, in this form, it is not particularly useful, because to evaluate the r.h.s one needs to know the exact density matrix $\hat{\rho}(t)$. 

Now comes the crucial step. Importantly, the operator $L^\dagger_{j} [  Q_a , L_{j} ]$ is local, because both the operator $L_j$ and the charge density $q_{a,j}$ have compact support. This, together with the assumption of slow losses, allows us to use the idea of local relaxation in the system. Namely, we expect that, under unitary evolution, the density matrix of a small subsystem quickly relaxes to a Generalized Gibbs Ensemble (GGE). Expectation values of local observables can then be evaluated with respect to the GGE density matrix,
\begin{align}\label{GGE}
    \hat{\rho}_{{\rm GGE}, \{ \langle Q_a \rangle \} } \, \propto \, e^{-\sum_{a} \beta_a Q_a},
\end{align}
where the Lagrange multipliers $\beta_a$ are fixed by the expectation values of the charges $\langle Q_a \rangle$, which must be equal to ${\rm tr} [  \hat{\rho}_{{\rm GGE}, \{ \langle Q_b \rangle \} } \, Q_a ]$. Evaluating the r.h.s of Eq.~(\ref{evoQ}) in the GGE density matrix leads to a closed evolution equation for the slow motion of the charges induced by the losses,
\begin{align}\label{evoQgge}
\frac{d}{dt} \langle Q_a \rangle = \Gamma \sum_{j=1 }^{L} {\rm Re} \, \expval{   L^\dagger_{j}  [ Q_a , L_{j} ] }_{{\rm GGE},  \{ \langle Q_b \rangle \} }.
\end{align}
It is this evolution equation that we study in great detail in this paper. For lattice HCB, the description is further simplified by specifying the form of the conserved charges $Q_a$. This is what we do next, by introducing the distribution of rapididites.

\subsection{Slow evolution of the rapidity distribution}
\label{subsec:rapidity}
Hard-core bosons can be mapped to free fermions by a Jordan-Wigner transformation, 
\begin{align}\label{JW}
\sigma^{+}_{j} = \prod_{i=1}^{j-1}(-1)^{c^\dagger_{i}c_i} \, c^\dagger_{j}, \qquad  \sigma^{-}_{j} = \prod_{i=1}^{j-1}(-1)^{c^\dagger_{i}c_i} \, c_{j} .
\end{align}
Here the operators $c^\dagger_j/c_j$ create/annihilate a fermion on site $j$. They satisfy the canonical anticommutation relations $\{c_{i},c^{\dagger}_{j}\}=\delta_{ij}$. Under the Jordan-Wigner mapping, the Hamiltonian (\ref{HCB}) becomes
\begin{equation}
    H_{\rm HCB} \, = \, - \frac{1}{2} \sum_{j =1}^{L} ( c^\dagger_{j} c_{j+1} + c^\dagger_{j+1} c_{j} ) .
\end{equation}
Moreover, the fermions satisfy antiperiodic (resp. periodic) boundary conditions if the number of particles $N$ in the system is even (resp. odd):
\begin{equation}
    \label{eq:bc}
    c_{L+1}^\dagger =  (-1)^{N-1} c_1^\dagger .  
\end{equation}
The Fourier modes are
\begin{equation}
    c^\dagger(k) = \frac{1}{\sqrt{L}} \sum_{j = 1}^{L} e^{i k j} c_j^\dagger 
\end{equation}
with $k \in \frac{2\pi}{L} (\mathbb{Z}+\frac{1}{2})$ if $N$ is even, and $k \in \frac{2\pi}{L} \mathbb{Z}$ if $N$ is odd. Either way the Hamiltonian reads
\begin{equation}
    \label{eq:Hfourier}
    H_{\rm HCB} \, = \,  \sum_k \varepsilon(k) c^\dagger(k) c(k) ,
\end{equation}
with $\varepsilon(k) = - \cos k$.

It is clear from the form (\ref{eq:Hfourier}) that any operator of the form
\begin{equation}
    Q[f]  = \sum_k f(k) c^\dagger(k) c(k) ,
\end{equation}
for any function $f(k)$, commutes with the Hamiltonian $H_{\rm HCB}$. Moreover these conserved charges also commute among themselves. Convenient choices for $f(k)$ are $\cos ( n k ) $ or $\sin (n k)$ for $n \in \mathbb{N}$, which leads to a hermitian basis set of charges, where each charge has a charge density that is compactly supported.

However, for the purposes of this paper, rather than to work with a specific choice of basis for the space of conserved charges $Q_a$ (or $Q[f]$), it is more convenient to work directly with the occupation number, or `rapidity distribution', 
\begin{equation}
    \rho(k) \, \underset{L \rightarrow \infty}{=} \, \left<  c^\dagger(k) c(k) \right> ,  \quad  \rho(k)  \, \in \, [0,1] .
\end{equation}
It is clear that if we know the rapidity distribution $\rho(k)$, then we also know the expectation values of any charge $Q[f]$, because $\left< Q[f] \right> =  \sum_k f(k) \rho(k)$.

Following Ref.~\cite{bouchoule_effect_2020}, we can turn the evolution equation for the slow motion of the charges (\ref{evoQ}) into an equation for the slow evolution of the rapidity distribution itself,
\begin{align}\label{rho_evo}
    \dot{\rho}(k)=-\Gamma F[\rho](k),
\end{align}
where the loss functional 
\begin{align}\label{FLoss}
    F[\rho](k)=\sum_{j=1}^{L} {\rm Re} \,  \expval{L^\dagger_j[L_j,c^\dagger(k)c(k)]}_{{\rm GGE}, \rho} ,
\end{align}
and the GGE density matrix itself is parameterized by the rapidity distribution $\rho(k)$. More precisely, the GGE density matrix is Gaussian for the fermions $c_j^\dagger, c_j$, and it is characterized by its two-point function $\expval{ c^\dagger(k) c(k') }_{{\rm GGE}, \rho} = \rho(k)  \delta_{k,k'}$. All higher-order correlations can be computed using Wick's theorem for fermionic operators.

The functional (\ref{FLoss}) is the central object of this paper. In the next section we compute it explicitly for K-body loss processes. For one- and two-boson loss processes we get simple closed expressions. For loss events involving larger numbers $K$ of bosons, we will see that we can express the loss functional as a small determinant, which follows from applying Wick's theorem to Eq.~(\ref{FLoss}).

\section{Deriving the loss functional}
\label{sec:lossfunctional}

In this section  we compute the functional $F[\rho](k)$ explicitly. Importantly, in our calculation we uncover a different structure depending on the parity of the number $K$ of bosons lost in each loss event, which can be traced back to the Jordan-Wigner string appearing in the mapping (\ref{JW}) to non-interacting fermions.

\subsection{One-body losses}\label{oneb}

For $K=1$ the Lindblad dissipators are $L_j=\sigma^-_j$. Using translational invariance, the loss functional (\ref{FLoss}) that we need to compute is
\begin{eqnarray}
    \label{eq:1loss_F}
\nonumber   && F[\rho](k) \, = \,  L \expval{ \sigma^+_1 [\sigma^-_1, c^\dagger(k) c(k) ] }_{{\rm GGE}, \rho} \\
\nonumber    && = \, L \expval{ \sigma^+_1 \sigma^-_1 c^\dagger(k) c(k) }_{{\rm GGE}, \rho} - L \expval{ \sigma^+_1 c^\dagger(k) c(k) \sigma^-_1 }_{{\rm GGE}, \rho} , \\
\end{eqnarray}
where $L$ is the length of the system and the loss operator acts on the site $j=1$. Both terms in the second line of (\ref{eq:1loss_F}) can be computed using the fact that the GGE is a gaussian state for the fermions, which allows us to use Wick's theorem. For the first term we have (using $c_1 = \frac{1}{\sqrt{L}} \sum_{q} e^{iq} \,c(q)$):
\begin{eqnarray}
    \label{eq:term1}
 \nonumber  && L \expval{ \sigma^+_1 \sigma^-_1 c^\dagger(k) c(k) }_{{\rm GGE}, \rho} \, = \, L \expval{ c^\dagger_1 c_1 c^\dagger(k) c(k) }_{{\rm GGE}, \rho} \\
\nonumber  &&  = \, \sum_{qq'} e^{i(q-q')} \expval{ c^\dagger(q') c(q) c^\dagger(k) c(k) }_{{\rm GGE}, \rho} \\
\nonumber  &&  = \, \sum_{qq'}  e^{i(q-q')} \left( \langle c^\dagger(q') c(q) \rangle  \langle c^\dagger(k) c(k) \rangle + \right. \\
\nonumber && \qquad \qquad \left. \langle c^\dagger(q') c(k) \rangle  \langle c(q) c^\dagger(k) \rangle \right) \\
&& = \, \langle N \rangle \rho(k) + \rho(k) (1-\rho(k)) .
\end{eqnarray}
The second term requires more care, because the operator $c^\dagger(k) c(k)$ is inserted between $\sigma_1^+$ and $\sigma_1^-$, and the latter change the parity of the number of particles in the system. The boundary conditions for the fermions are modified according to Eq.~(\ref{eq:bc}). Thus we need to relate the Fourier modes of the fermions with periodic boundary conditions to the ones with anti-periodic boundary conditions. For conciseness, let us introduce the two corresponding sets of momenta,
\begin{align}
    &Q^{\rm p} = \frac{2\pi}{L} \times \{1,2 , \dots, L  \}  , \\   
    &Q^{\rm ap} =  \frac{2\pi}{L} \times  \{ \frac{1}{2} , \frac{3}{2},  \dots,  L- \frac{1}{2} \}  .
\end{align}
Then we have the following identities,
\begin{align}\label{parity_relat}
(k \in Q^{\rm p} ) \qquad c(k)=\dfrac{i}{L} \sum_{q \in Q^{\rm ap}  } \dfrac{e^{i(q-k)/2}}{\sin((q-k)/2)}\, c(q), \\ \qquad
(k \in Q^{\rm ap} ) \qquad c(k)=\dfrac{i}{L} \sum_{q \in Q^{\rm p}  } \dfrac{e^{i(q-k)/2}}{\sin((q-k)/2)}\, c(q) .
\end{align}
We can insert them into the second term of Eq.~(\ref{eq:1loss_F}), which leads to
\begin{align}
\expval{ \sigma^+_1  c^\dagger(k) c(k) \sigma^-_1}=\dfrac{1}{L^{2}} \sum_{q,q'} \dfrac{e^{i(q-q')/2} \, \expval{ c^{\dagger}_1 c^{\dagger}(q')\, c(q) c_1 } }{\sin((q-k)/2)\sin((q '-k)/2)}.  
\end{align}
This correctly implements the change of boundaries of the fermions. Next we can apply Wick's theorem to evaluate the four-fermion correlator $\expval{ c^{\dagger}_1 c^{\dagger}(q')\, c(q) c_1 }$. This leads to
\begin{align}
  & 
  L \expval{\sigma^+_1  c^\dagger(k) c(k) \sigma^-_1} = \nonumber \\
  &\frac{\expval{N}}{L^{2}}\sum_{q} \dfrac{\rho(q)}{\sin^{2}(\frac{q-k}{2})}-\left(\frac{1}{L}\sum_{q}\cot(\frac{q-k}{2}) \, \rho(q)\right)^{2}- \dfrac{\expval{N}^{2}}{L^{2}}.
\end{align}
The first term in the above equation has a pole of order 2 at $q=k+2 \pi \mathbb{Z}$, and it is convenient to reduce its degree using the identity $\sum_q 1/\sin^{2}(\frac{q-k}{2})=L^2$. This leads to the equivalent expression
\begin{align}
    \label{eq:term2}
    L \expval{\sigma^+_1  c^\dagger(k) c(k) \sigma^-_1}&=\frac{\expval{N}}{L^{2}}\sum_{q} \dfrac{\rho(q)-\rho(k)}{\sin^{2}(\frac{q-k}{2})}+\expval{N}\rho(k)\nonumber \\ &-\left(\frac{1}{L}\sum_{p}\cot(\frac{p-k}{2}) \, \rho(p)\right)^{2}- \dfrac{\expval{N}^{2}}{L^{2}}.
\end{align}
Putting the two terms (\ref{eq:term1})-(\ref{eq:term2}) together and taking the thermodynamic limit $L \rightarrow \infty$, we arrive at the following form of the one-body loss functional
\begin{align}\label{1loss_Fres}
    F[\rho](k)&=\rho(k)-\rho^2(k)+\left(\fint^\pi_{-\pi} \frac{dp}{2\pi} \, \cot(\frac{k-p}{2}) \, \rho(p)\right)^{2}\nonumber \\
    &+n\left(n+\fint^\pi_{-\pi} \frac{dq}{2\pi} \, \dfrac{\rho(k)-\rho(q)}{\sin^{2}(\frac{k-q}{2})}\right),
\end{align}
where $n=\expval{N}/L$ is the density of particle and $\fint$ means the Cauchy principal value of the integral. This is our main result for one-body losses. It is similar to (but different from) the formula given in Ref. \cite{bouchoule_effect_2020} for the Tonks-Girardeau gas in the continuum. Notice that the functional is non-linear in $\rho(k)$, and also non-local in rapidity space.




\subsection{Two-body losses}\label{2b}
For $K=2$, the dissipators are $L_j= \sigma^-_j \sigma^-_{j+1}$. 
Under the Jordan-Wigner mapping they become
\begin{ceqn}
\begin{align}
    \label{eq:dissipK2}
L_{j}= \sigma^-_{j} \sigma^-_{j+1}=   c_{j} (-1)^{c^{\dagger}_{j}c_{j}}  c_{j+1} = - c_j c_{j+1} .
\end{align}
\end{ceqn}
Then, to compute the functional $F$, we simply need to insert the dissipator (\ref{eq:dissipK2}) in the definition \eqref{FLoss},
\begin{ceqn}
\begin{align}
&F[\rho](k)=\sum_{j}\expval{\sigma^{+}_{j+1}\sigma^{+}_{j}[\sigma^-_{j}\sigma^-_{j+1},c^\dagger(k) c(k)]}\nonumber \\
&=\sum_{q,q',p,p'}\frac{e^{i(2p'+p-2q-q')}}{M}\expval{c^{\dagger}(q)c^{\dagger}(q')[c(p)c(p'),c^{\dagger}(k)c(k)]}.
\end{align}
\end{ceqn}
Expanding the commutator in the braket leads to two terms
\begin{align}\label{deuxtermes}
    &\expval{c^{\dagger}(q)c^{\dagger}(q')[c(p)c(p'),c^{\dagger}(k)c(k)]}\nonumber\\
    &=\expval{c^{\dagger}(q)c^{\dagger}(q')c(p)c(p')c^{\dagger}(k)c(k)}\nonumber\\
    &-\expval{c^{\dagger}(q)c^{\dagger}(q')c^{\dagger}(k)c(k)c(p)c(p')}.
\end{align}
The first term can be expressed as
\begin{align}\label{secterm}
    &\expval{c^{\dagger}(q)c^{\dagger}(q')c(p)c(p')c^{\dagger}(k)c(k)}\nonumber\\
    &=\expval{c^{\dagger}(q)c^{\dagger}(q')c^{\dagger}(k)c(k)c(p)c(p')}\nonumber \\
    &+\delta_{p'k}\expval{c^\dagger(q) c^\dagger(q') c(p) c(k)}-\delta_{pk} \expval{c^\dagger(q) c^\dagger(q') c(p') c(k)}.
\end{align}
The second term is the expectation value of $c^{\dagger}(k)c(k)$ in a state where two atoms have been removed, the parity of the initial number of particle is then unchanged. Hence, in contrast to the $K=1$ case treated in the previous subsection, here the parity of the number of atoms does not change.

One can then apply Wick's theorem, and take the thermodynamic limit to obtain the loss functional, 
\begin{align}\label{2loss_F}
F[\rho](k)&=\dfrac{2}{\pi} \int^\pi_{-\pi} dq \,\sin^{2}\left (\frac{k-q}{2} \right) \; \rho(q) \, \rho(k).
\end{align}
This is our main result for two-body losses. That functional presents some similarities with the loss functional found in the relation (5) of Ref.~\cite{rossini_strong_2021}.

We now generalise this calculation to the case of $K$-body losses for $K$ an arbitrary even integer.

\subsection{$K$-body losses with K even}\label{2c}
In this subsection we show that is possible to find a closed formula for the loss functional defined in~\eqref{FLoss} where the Lindblad operator is given by $L_j=\sigma_j\sigma_{j+1}\dots \sigma_{j+K-1}$. Taking the Fourier transform of $L_j$ and $L^\dagger_j$ in~\eqref{FLoss}, the loss functional reads 
\begin{align}\label{general_floss}
    F^{\rm even}[\rho](k)=&\frac{1}{L^{K-1}}\sum_{\substack{q_1,\dots,q_K\\ q'_1,\dots,q'_K}}\exp{i\sum^K_{l=1}(q_l-q'_l)l} \times\nonumber \\
    &\expval{c^\dagger(q'_K) \dots c^\dagger(q'_1)[c(q_1)\dots c(q_K),c^\dagger(k)c(k)]}.
\end{align}
As mentionned in \ref{2b}, in the case of even $K$-body losses the commutator in~\eqref{general_floss} reduces to $K$ terms 
\begin{align}
    &\expval{c^\dagger(q'_K) \dots c^\dagger(q'_1)[c(q_1)\dots c(q_K),c^\dagger(k)c(k)]}\nonumber \\
    &=\delta_{kq_K}\expval{c^\dagger(q'_K) \dots c^\dagger(q'_1)c(q_1)\dots c(q_{K-1})c(k)}\nonumber \\
    &-\delta_{kq_{(K-1)}}\expval{c^\dagger(q'_K) \dots c^\dagger(q'_1)c(q_1)\dots c(q_{K-2}) c(q_{K})c(k)}\nonumber\\
    &+\dots
\end{align}
and applying Wick's theorem on each terms leads to a product of $K$ terms of the form $\expval{c^\dagger(q')c(q)}$. Using the property $\expval{c^\dagger(q')c(q)}=\delta_{qq'}\, \rho(q)$ and taking the thermodynamic limit, the loss functional in~\eqref{general_floss} can be expressed as a sum of $K$ terms each consisting of a $K$ by $K$ matrix determinant. Let us introduce the $K \times K$ matrix $A^{(j)}_{[\rho]}$ with matrix elements
\begin{align}\label{matriceA}
     [A^{(j)}_{[\rho]} ]_{ab} = \left\lbrace 
\begin{array}{ll}
\frac{1}{2\pi}\int^\pi_{-\pi}dq \, e^{i(b-a)q} \, \rho(q) \quad {\rm if} \; b \neq j \\
e^{i(b-a)k} \, \rho(k)\quad {\rm if} \; b = j ,
\end{array}
\right.
\end{align}
for indices $a,b= 1, \dots , K$. The superscript $j$ indicates which column depends on the rapidity $k$. Apart from the $j^{\rm th}$ column, the matrix essentially contains Fourier transforms of the rapidity distribution $\rho(k)$. The loss functional then reduces to 
\begin{align}\label{Kfunc_even}
F^{\rm even}[\rho](k)= \sum^{K}_{j=1} \det(A^{(j)}_{[\rho]}) .
\end{align} 
The relation~\eqref{Kfunc_even} is another fundamental result of this paper.

\begin{figure*}[htb]
    \centering
    \includegraphics[scale=0.565]{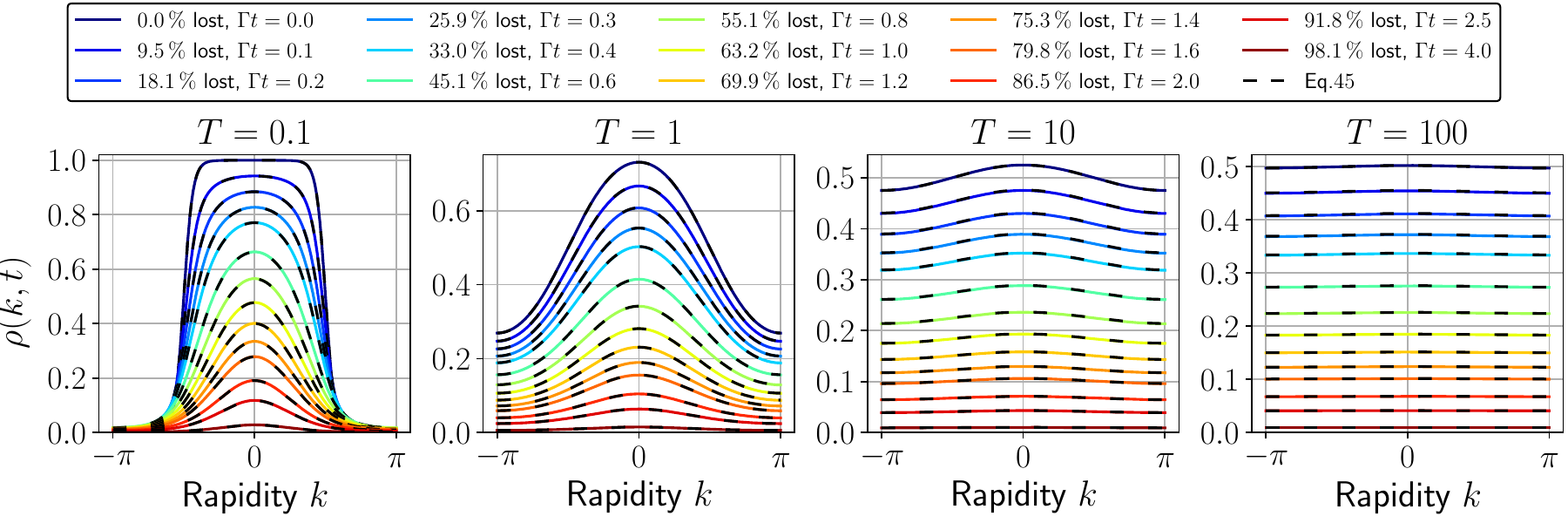}
    \caption{ {\color{black} Effect of one-body losses on different rapidity distributions. The initial distribution is a thermal (Fermi-Dirac) distribution $\rho(k)=(1+\exp(-\cos(k)/T))^{-1}$, and we show the results for different initial temperatures $T=0.1,1.0,10.0,100.0$. The time evolution is obtained by solving Eq.~\eqref{rho_evo} with the loss functional (\ref{1loss_Fres}) with the Runge-Kutta method. Notice that the solution of Eq.~\eqref{rho_evo} depends on the loss rate $\Gamma$ only through the dimensionless time $\Gamma t$. The black dashed lines correspond to formula (\ref{result_rho}) and show a perfect agreement with the numerical solution of the evolution equation. } }
    \label{1b_kdistrib}
\end{figure*}

\subsection{$K$-body losses with K odd}\label{2d}
In this last subsection we investigate the case of losses process with an odd number $K$ of lost atoms. The reasoning is similar to the one we developped in the previous subsection, however like in the $K=1$ case, we need to be careful about the change of boundary conditions for the fermions. We start by taking the Fourier transform of the Lindblad operators in the definition~\eqref{FLoss}, which leads to the relation~\eqref{general_floss}. The commutator in the loss functional gives two terms
\begin{align}\label{commut}
    &\expval{c^\dagger(q'_K) \dots c^\dagger(q'_1)[c(q_1)\dots c(q_K),c^\dagger(k)c(k)]}\nonumber \\
    &=\expval{c^\dagger(q'_K) \dots c^\dagger(q'_1)c(q_1)\dots c(q_K)c^\dagger(k)c(k)}\nonumber\\
    &-\expval{c^\dagger(q'_K) \dots c^\dagger(q'_1)c^\dagger(k)c(k)c(q_1)\dots c(q_K)}.
\end{align}
As we have already discussed in the subsection~\ref{oneb}, the first term is the expectation value of $c^\dagger(k)c(k)$ in a state where the initial number of particles is preserved. However the second term corresponds to the expectation value of $c^\dagger(k)c(k)$ in a state where $K$ atoms have been removed. Since $K$ is an odd number, the parity of the number of particle is changed and one needs to use the relations~\eqref{parity_relat} to express $c^\dagger(k)c(k)$ in the appropriate parity sector. Inserting the relations~\eqref{parity_relat} in the second term, one has
\begin{align}\label{commute_1}
    &\expval{c^\dagger(q'_K) \dots c^\dagger(q'_1)c^\dagger(k)c(k)c(q_1)\dots c(q_K)} =\nonumber \\
    &\sum_{q,q'} \dfrac{e^{i(q-q')/2} \, \expval{ c^\dagger(q'_K) \dots c^\dagger(q'_1)c^\dagger(q')c(q)c(q_1)\dots c(q_K) } }{L^{2}\sin((q-k)/2)\sin((q '-k)/2)}. 
\end{align}
Before using Wick's theorem on the above formula, one can notice that the first term in the right-hand side of~\eqref{commut} can be written as
\begin{align}\label{commute_2}
    &\expval{c^\dagger(q'_K) \dots c^\dagger(q'_1)c(q_1)\dots c(q_K)c^\dagger(k)c(k)}\nonumber \\
    &=\expval{c^\dagger(q'_K) \dots c^\dagger(q'_1)c^\dagger(k)c(k)c(q_1)\dots c(q_K)}\nonumber \\
    &+\delta_{kq_K}\expval{c^\dagger(q'_K) \dots c^\dagger(q'_1)c(q_1)\dots c(q_{K-1})c(k)}\nonumber \\
    &-\delta_{kq_{(K-1)}}\expval{c^\dagger(q'_K) \dots c^\dagger(q'_1)c(q_1)\dots c(q_{K-2}) c(q_{K})c(k)}\nonumber\\
    &+\dots,
\end{align}
where we used the anti-commutation relation for the fermionic operators. As we proceed in the previous subsection, the Wick's contractions of~\eqref{commute_1} and of the first term in the right-hand side of~\eqref{commute_2} can be written as two determinants of two matrices $B$ and $C$.    
The matrices $B$ and $C$ are $(K+1) \times (K+1)$ hermitian matrices and their matrix elements depend on the Fourier and Hilbert transforms \cite{butzer_fourier_1971} of $\rho(k)$  
\begin{align}\label{matriceB}
     [B_{[\rho]} ]_{ab} = \left\lbrace 
\begin{array}{ll}
\frac{1}{2 \pi} \int^{\pi}_{-\pi}dq \, e^{i(b-a)q}\, \rho(q) \; \; {\rm if} \;a, \, b < K+1 \\
e^{-iak} \, \rho(k) \;\; {\rm if} \; b = K+1 \\
0 \; {\rm if} \; a=b=K+1 ,
\end{array}
\right.
\end{align}
\begin{align}\label{matriceC}
     [C_{[\rho]}]_{ab} = \left\lbrace 
\begin{array}{ll}
\frac{1}{2 \pi} \int^{\pi}_{-\pi}dq \, e^{i(b-a)q}\, \rho(q) \; {\rm if} \;a, \, b < K+1 \\
\frac{1}{2\pi} \fint^\pi_{-\pi}dq \, e^{-i(a-1)q}\rho(q)(\cot(\dfrac{k-q}{2})+i ) \\ {\rm if} \; b = K+1 \\
\frac{1}{2\pi}\fint^{\pi}_{-\pi}dq \, \dfrac{\rho(q)-\rho(k)}{\sin^2(\dfrac{k-q}{2})}\; {\rm if} \; a=b=K+1 .
\end{array}
\right.
\end{align}
The loss functional for odd $K$ takes the final form
\begin{align}\label{Kfunc}
F^{\rm odd}_{K}[\rho](k)&=\left( \sum^{K}_{j=1} \det(A^{(j)}_{[\rho]}) \right)\nonumber \\
&+\left[ \det(B_{[\rho]})-\det(C_{[\rho]})\right].
\end{align}

It is possible to write a general expression valid both for even and odd $K$ by introducing the factor $\dfrac{1-(-1)^{K}}{2}$ which vanishes for $K$ even, so that our final result, valid in all cases, reads 
\begin{align}\label{Kfunc2}
F_{K}[\rho](k)&=\left( \sum^{K}_{j=1} \det(A^{(j)}_{[\rho]} ) \right)\nonumber \\
&+ \dfrac{1-(-1)^{K}}{2} \left[ \det(B_{[\rho]})-\det(C_{[\rho]})\right].
\end{align}

\section{Evolution of the rapidity distribution in a homogeneous gas}
\label{sec:homogeneous}

Having established the general form of the loss functional $F[\rho]$ for $K$-body losses,  Eqs.~(\ref{1loss_Fres})-(\ref{2loss_F})-(\ref{Kfunc_even})-(\ref{Kfunc})-(\ref{Kfunc2}), we now turn to the time evolution of the rapidity distribution. We solve the evolution equation
\begin{align}
    \label{eq:evol_rhok}
    \dot{\rho}(k)=-\Gamma F[\rho](k)
\end{align}
numerically (and analytically for the special case $K=1$), and we focus in particular on the time evolution of the atom density $n = \int_{-\pi}^\pi \rho(k) \frac{dk}{2\pi}$.

\subsection{Results for $K=1$}

For one-body losses the atom density always decays exponentially,
\begin{equation}
    \label{eq:evol_n_K1}
    n(t)=n(0) e^{-\Gamma t} .
\end{equation}
This simply follows from Eq.~(\ref{evoQ}) applied to the total particle number $Q_a = N$ and to $L_j = \sigma^-_j$: it gives $\langle \dot{N} \rangle = - \Gamma  \langle N  \rangle $, which implies Eq.~(\ref{eq:evol_n_K1}) for the atom density $n = N/L$.


It turns out the evolution equation (\ref{eq:evol_rhok}) for the rapidity distribution can be solved exactly for the loss functional for $K=1$ (see Eq.~(\ref{1loss_Fres})). The solution is derived in Appendix~\ref{appendix A}; it reads
\begin{align}\label{result_rho}
	\rho(t,k)=n_0e^{-\Gamma t}\Re \left(\dfrac{\tanh(n_0(e^{-\Gamma t}-1))+\frac{i}{ n_0} \, I(t,k)}{1+\frac{i}{ n_0}\tanh(n_0(e^{-\Gamma t}-1)) \; I(t,k)}\right) ,
\end{align}
where $n_0=n(0)$ is the initial atom density, $\rho_0(k)=\rho(t=0,k)$ is the initial rapidity distribution, and $I(t,k)$ is the integral 
\begin{align}\label{Q(k)}
    I(t,k)= \int^{\pi}_{-\pi} \frac{dq}{2\pi} \dfrac{\rho_0(q)}{\tan(\frac{k-q}{2}+in_0(1-e^{-\Gamma t}))}.
\end{align}
A related expression for the continuous Tonks-Girardeau gas with one-body losses was obtained in Ref.~\cite{bouchoule_effect_2020}.


In Fig.~(\ref{1b_kdistrib}) we show the evolution of the rapidity distribution from thermal initial states at different temperatures. We display the analytical result (\ref{result_rho}), as well as the numerical solution of Eq.~(\ref{eq:evol_rhok}) obtained from the Runge-Kutta method; they are in perfect agreement, as they should.

\begin{figure}[t]
    \centering
    \includegraphics[scale=0.565]{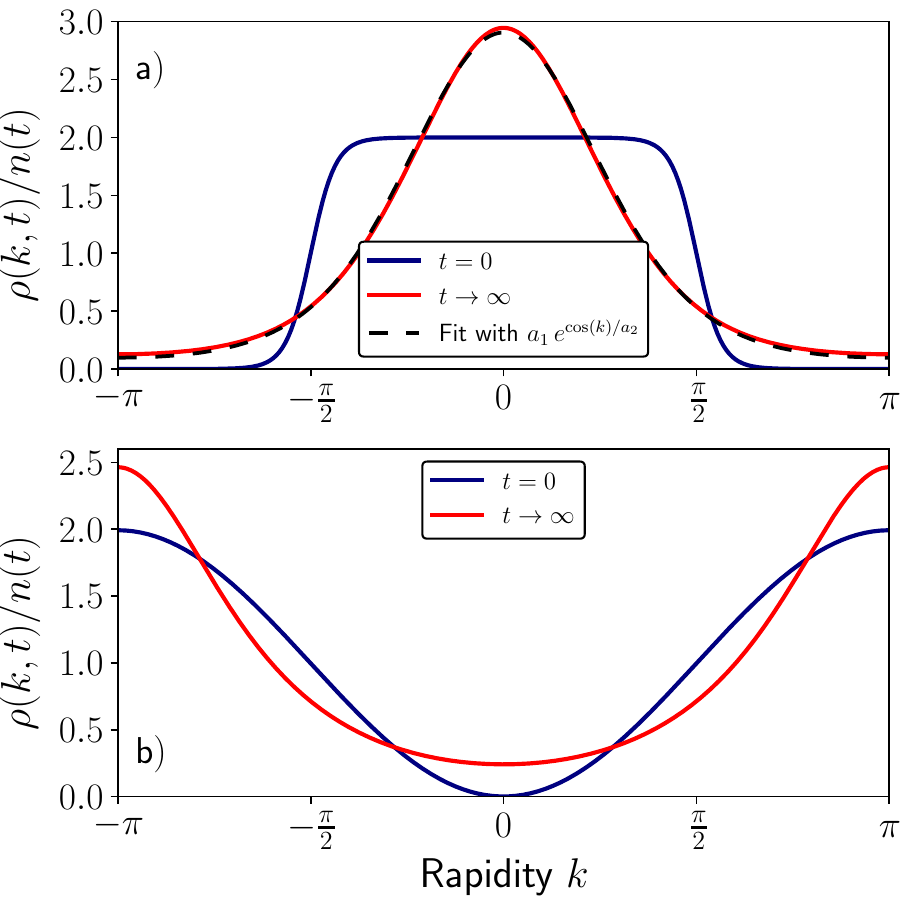}
    \caption{{\color{black} Long time behavior of different rapidity distributions rescaled by the corresponding density. a) The initial distribution (blue curve) is a Fermi-Dirac distributions $\rho_0(k)=(1+\exp(-\cos(k)/T))^{-1}$ with $T$=0.1. The red curve is the distribution at long time according to Eq.~\eqref{result_rho}. The black dashed line is a numerical fit with a Boltzmann distribution $a_1 \, \exp (\cos(k)/a_2)$. b) The initial distribution (blue curve) is a cosine function $\rho_0(k)=(1-\cos(sk))/2$. The red curve is the distribution for $t \rightarrow \infty$. Here it is clear that the distribution does not become thermal.}}
    \label{longtime_cos}
\end{figure}

We see that loss processes spread the distribution in rapidity space. In the limit of large temperature for the initial state, the rapidity distribution is flat, and remains flat at all times. For smaller initial temperatures, it evolves into a bell-shape distribution at late times, which is close to a Boltzmann distribution $\rho(k) \propto \exp [ \cos (k)/T ]$ with a density going to zero according to Eq.~(\ref{eq:evol_n_K1}). This is further illustrated in Fig.~\ref{longtime_cos}.(a), where we plot the ratio $\rho(k,t)/n(t)$ in the limit $t\rightarrow \infty$, and fit the result with a Boltzmann distribution. The agreement is very good, even though it is clear from the exact formulas~(\ref{result_rho})-(\ref{Q(k)}) that the distribution $\rho(k,t)$ never becomes exactly thermal, even at infinite time.

To find a more striking signature of the fact that the system never goes to a low-density thermal distribution, we consider the case of an oscillating initial rapidity distribution $\rho_0(k)=(1-\cos(sk))/2$, where $s$ is an integer. In that case the long-time limit of the integral~\eqref{Q(k)} can be evaluated analytically, $ \lim\limits_{t \rightarrow \infty} I(t,k) = (1-e^{-2sn_0} e^{iks})/2$, and when injected in Eq.~(\ref{result_rho}) it leads to a late-time rapidity distribution of the form
\begin{align}\label{rho_ini_cos2}
\frac{\rho(k,t)}{n(t)}  \, \underset{t\rightarrow \infty}{=}   \, \frac{\alpha+\beta \, \cos(sk)}{\gamma+\delta \, \cos(sk)},
\end{align}
where the coefficients $\alpha, \beta, \gamma, \delta$ depend on the initial density $n_0$ and on the integer $s$. Thus, even at long time, the rescaled rapidity distribution is sensitive to the structure of the inital distribution (see also Fig.\ref{longtime_cos}.(b)). We conclude that in general the rapidity distribution does not go to a low-density thermal distribution at long times.

\begin{figure*}[htb!]
    \centering
    \includegraphics[scale=0.565]{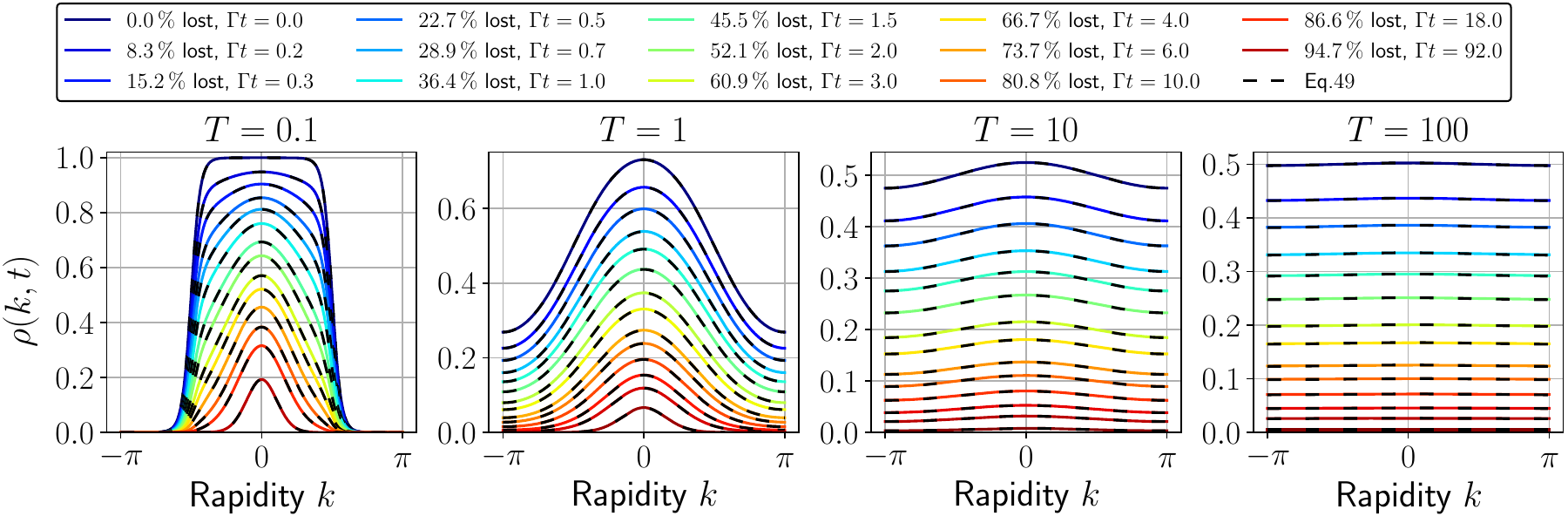}
    \caption{{\color{black} Effect of two-body losses on the rapidity distribution. The initial rapidity distribution $\rho(k,0)=1/(1+\exp(-\cos(k)/T))$ is chosen with $T=0.1, 1, 10, 100$ from left to right. The black dashed curve is the analytic solution (see Eq.~\eqref{Eq:rhokt:2bl}). After several loss events the rapidity distribution takes the form of a gaussian centered at $k=0$. As in the one-body case, if initially the rapidity distribution is flat then it remains flat under lossy evolution.}}
    \label{2bmany}
\end{figure*}

\subsection{Results for $K=2$}

We now consider the time evolution equation for the rapidity distribution for the $K=2$ case, which is characterized by the functional given by Eq.~\eqref{2loss_F}. For simplicity, here we focus on initial rapidity distributions $\rho_0(k)$ that are symmetric under reflection  $k\rightarrow -k$. Since the master equation is also invariant under $k \rightarrow-k$, this property is conserved throughout the entire evolution. Then Eq.~\eqref{2loss_F} simplifies to the following expression,
\begin{equation}
F[\rho] = 2 \, \rho(k) \, n(t) - \dfrac{1}{\pi} 
\cos{(k)} \rho(k) \int^\pi_{-\pi} dq \, \cos{(q)} \; \rho(q).
\label{Eq:2bl:symm}
\end{equation}
Eq.~\eqref{Eq:2bl:symm} highlights the two distinct contributions to the time evolution of $\rho(k,t)$: the first term in the right-hand side represents a mean-field contribution, as it does not introduce any structure in rapidity space, while the second term is responsible for generating quantum correlations and, consequently, introducing structure in $k-$space.

After some algebra presented in App.~\ref{App:der:rhokt:tbl}, one can derive an exact (implicit) expression for the rapidity distribution at all times:
\begin{align}
 \nonumber   & \rho(k,t) =  \rho_0(k) 
    \\ &   \times \exp{ -2 \Gamma  \int_0^t  \left( 1 -  \sigma_0 \cos{(k)} \sqrt{ 1 + \frac{\partial_{\tau} n(\tau)}{2 \Gamma \, n(\tau)^2}  } \right) n(\tau ) d\tau },
    \label{Eq:rhokt:2bl}
\end{align}
where $ \sigma_0 = \text{sgn} (\int_{-\pi}^{\pi} \cos{(k)} \rho_0(k) dk)$, with $\text{sgn} (x) = \pm 1$ the sign function. Eq.~\eqref{Eq:rhokt:2bl} shows that $\rho(k,t)$ is entirely determined by $\rho_0(k)$ and $n(t)$. Notably, it reveals that rapidities are distributed according to a cosine law, with the $k=0$ mode that has the longest lifetime.

\begin{figure}[t]
    \centering
    \includegraphics[scale=0.35]{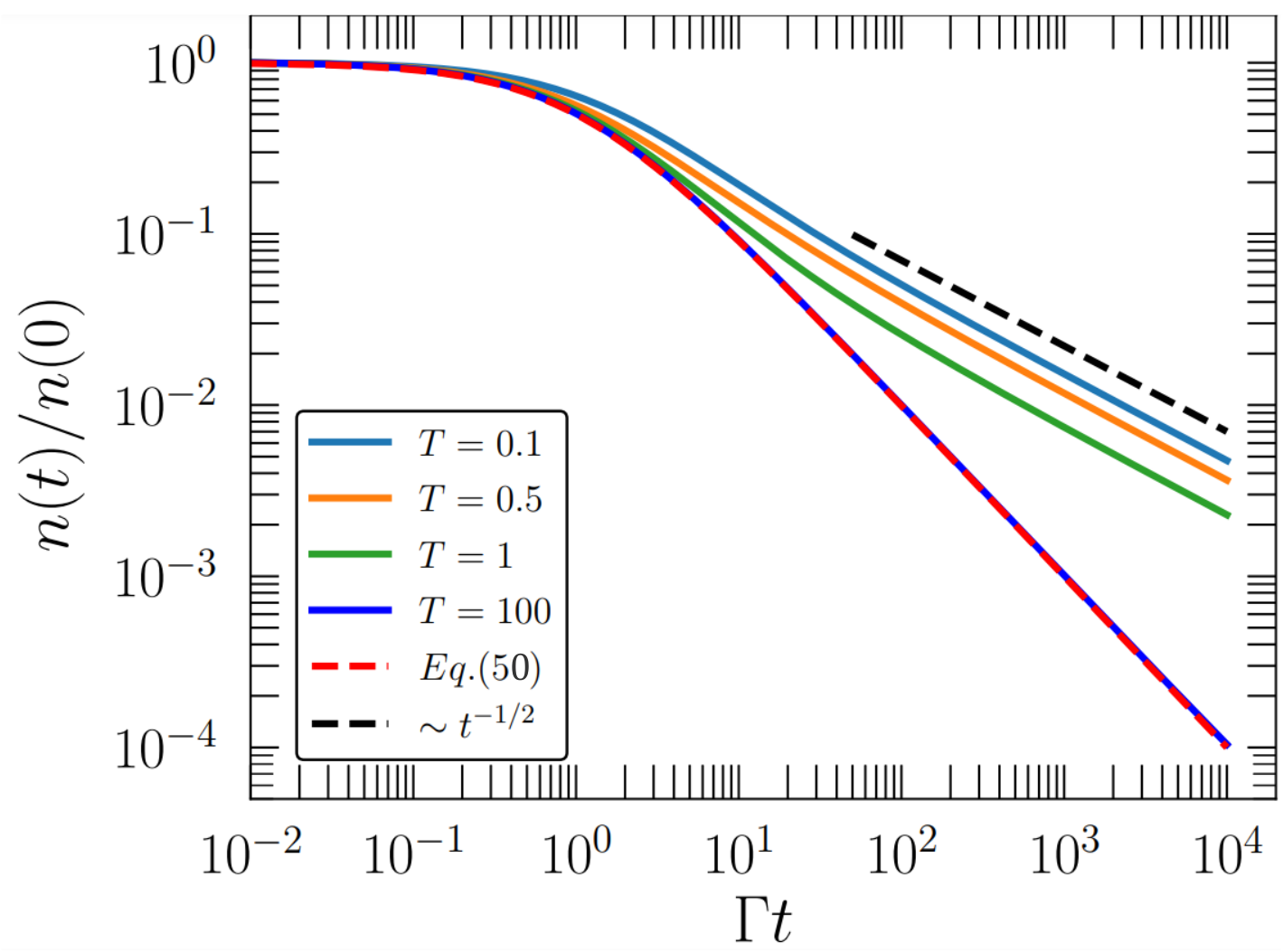}
    \caption{{\color{black} The mean density under two-body losses for different thermal initial rapidity distributions $\rho(k,0)=1/(1+\exp(-\cos(k)/T))$. Colored curves are obtained by solving numerically the time evolution equation ~\eqref{rho_evo} for the loss functional~\eqref{2loss_F} with the Runge-Kutta method. The red dashed line is the mean density associated to an initial rapidity distribution which is flat (i.e. infinite temperature), see Eq.~\eqref{n(k)K}. The dashed line is the expected long-time behavior $\sim t^{-1/2}$, see Eq.\eqref{longtime_2bdens}.}}
    \label{density_2body}
\end{figure}

In Fig.~\ref{2bmany} we show the evolution of the rapidity distribution for initial thermal states at different temperatures, and in Fig.~\ref{density_2body} we show the corresponding evolution of the mean atom density $n(t)$. It appears that, except for an initial infinite temperature state, the mean density always decays as $n(t) \propto 1/\sqrt{t}$ at very long times, while the density decreases as $1/t$ for an infinite temperature. These two behaviors follow from Eqs.~\eqref{Eq:2bl:symm}-(\ref{Eq:rhokt:2bl}), as we now explain.

Eq.~\eqref{Eq:2bl:symm} reveals that initial rapidity distributions that have a vanishing first Fourier mode $\int_{-\pi}^\pi dq \, \cos(q) \rho(q)$ always follow the exact same dynamics as the mean density, namely
\begin{equation}
    n(t) = \dfrac{1}{1 + 2 n(0) \Gamma t},
    \label{Eq:tbl:mf}    
\end{equation}
characterized by a long-term decay as $\sim 1/t$. This power law is then always found for initial rapidity distributions with vanishing first Fourier mode, including infinite temperature states.

\begin{figure}[h]
    \centering
    \includegraphics[scale=0.565]{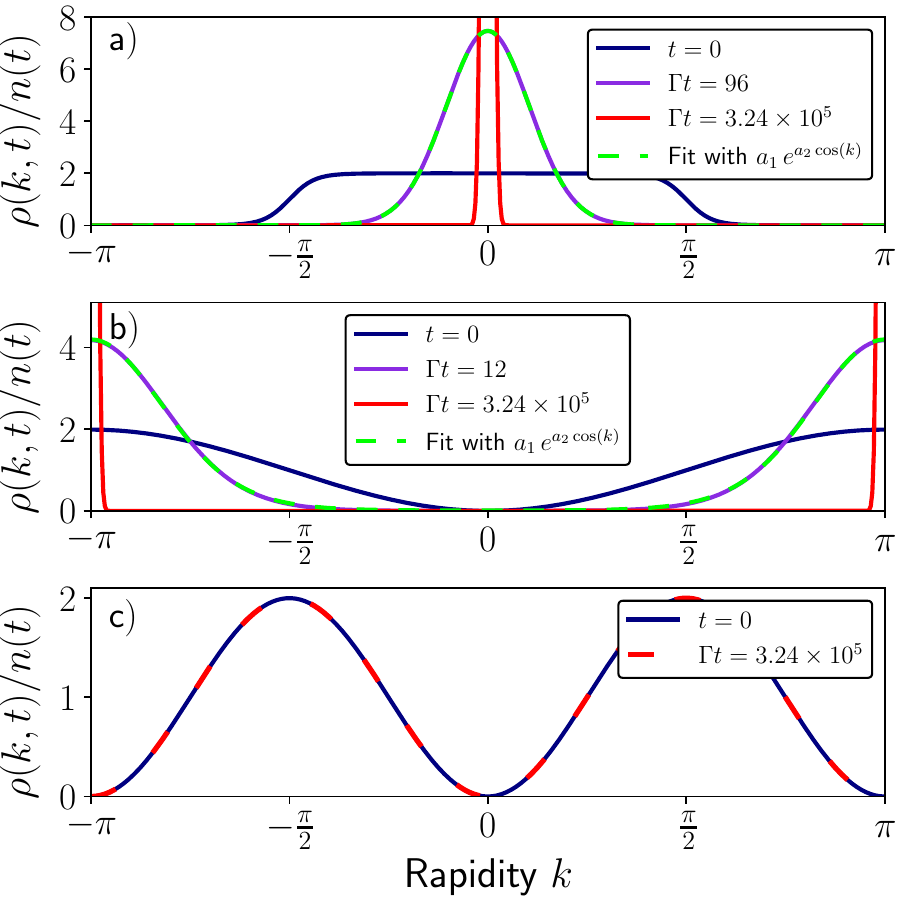}
    \caption{{\color{black}Long time behavior of different rapidity distributions rescaled by the corresponding density $n(t)$ under two-body losses. In a), b) and c), the blue curve represents the initial rapidity distribution while the red curve is the rapidity distribution at long time. The violet curve is the rescaled rapidity distribution at an intermediate time. a) The initial distribution is a Fermi-Dirac distributions $\rho_0
    (k)=(1+\exp(-\cos(k)/T))^{-1}$ with $T$=0.1. The green dashed curve is a fit with a Boltzmann distribution $a_1 \, \exp(a_2 \cos(k))$ where $a_2$ is positive. b) The initial rapidity distribution is $(1-\cos(k))/2$ which has a non-vanishing first Fourier mode. The green dashed curve is a fit with a Boltzmann distribution $a_1 \, \exp{a_2 \cos(k)}$ where $a_2$ is negative. c) The initial rapididty distribution is $(1-\cos(2k))/2$  which has no first Fourier mode.}}
    \label{figlongtime2}
\end{figure}

\begin{figure*}[htb]
    \centering
    \includegraphics[scale=0.565]{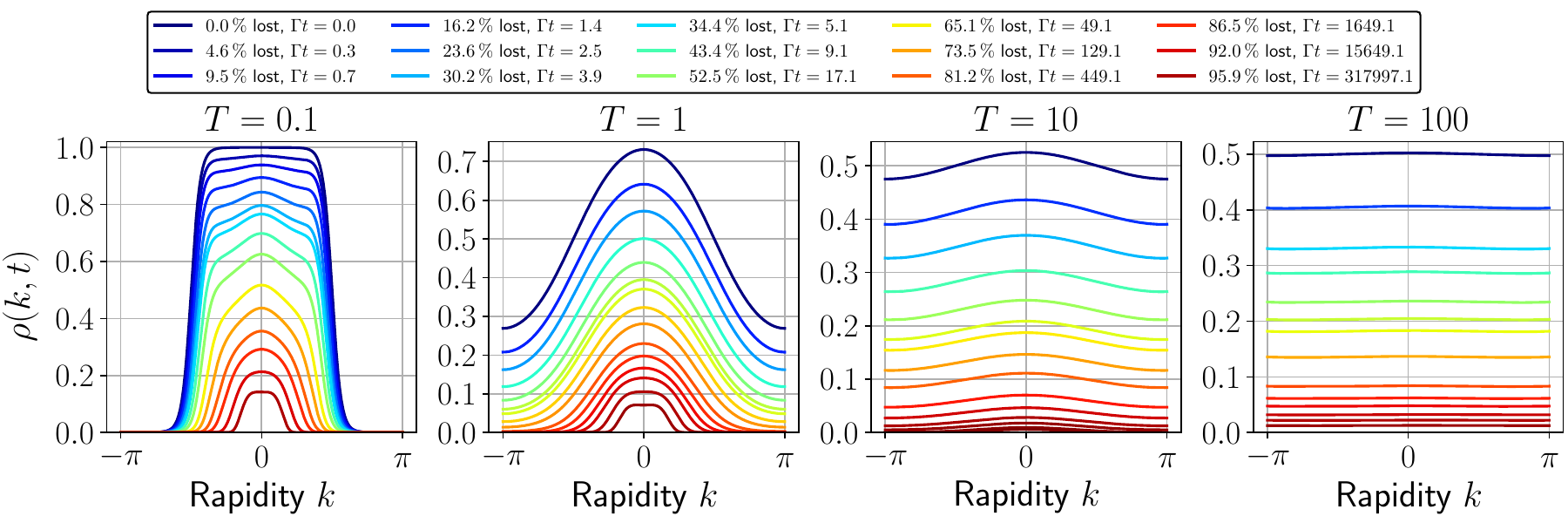}
    \caption{{\color{black} Evolution of the rapidity distribution under three-body losses. We numerically solve~\eqref{rho_evo} for the loss functional $F[\rho](k)$ corresponding to $K=3$, using the Runge-Kutta method. The initial rapidity distribution is the Fermi-Dirac distribution $\rho(k,0)=1/(1+\exp(-\cos(k)/T))$ at half filling and for different temperatures $T=0.10,1.0,100.0$ (from left to right).}}
    \label{3b_kdis}
\end{figure*}

In contrast, initial rapidity distributions that have a non-vanishing first Fourier mode decay as $\sim 1/\sqrt{t}$ at long times. This can be understood by looking at the long time limit of Eq.~\eqref{Eq:rhokt:2bl}. Let us introduce the two time-dependent functions $g(t)=\int_0^t n(\tau ) \,d\tau$ and $f(t)=\int_0^t\sqrt{ 1 + \frac{\partial_{\tau} n(\tau)}{2 \Gamma \, n(\tau)^2}  }  n(\tau ) d\tau $. Numerically we observe that $ | \partial_{\tau} n(\tau) | \ll 2 \Gamma n(\tau)^2$ at long times, as soon as the initial rapidity distributions has a non-vaishing first Fourier mode. Then, expanding at first order in $ | \partial_{\tau} n(\tau) | / ( 2 \Gamma n(\tau)^2) $, $f(t)$ becomes
\begin{align}\label{fandg}
    f(t) \simeq g(t)+\frac{1}{4\Gamma }\ln( \dfrac{n(t)}{n(0)}),
\end{align}
implying that, at large $t$, the difference between $f(t)$ and $g(t)$ grows as $\ln(t)$. Integrating Eq.~\eqref{Eq:rhokt:2bl} over $k$ leads to the mean density
\begin{align}
    n(t)=\frac{ e^{ -2 \Gamma  g(t)}}{2 \pi}\int^{\pi}_{-\pi}dk \, \rho_0(k) \,   e^{ 2 \Gamma \sigma_0 \cos{(k)} f(t)}.
\end{align}
Since the function $f(t)$ diverges at large $t$, the latter integral can be evaluated by the saddle-point approximation; we denote $k^*_{\sigma_0}$ the saddle point: $k^*_{\sigma_0} = 0$ if $\sigma_0 = +1$ and $k^*_{\sigma_0} = \pi$ if $\sigma_0 = -1$. We are thus left with:
\begin{align}
    \int^{\pi}_{-\pi}dk \, \rho_0(k) \,   & e^{ 2 \Gamma \sigma_0 \cos{(k)} f(t)}   \nonumber \\
    & \underset{t \rightarrow \infty}{\simeq} e^{ 2\Gamma \,f(t)}\int^{\infty}_{-\infty}dk \, \rho_0(k) \,   e^{ - \Gamma \, k^2 f(t)}\nonumber \\
    &=\sqrt{\frac{\pi}{\Gamma f(t)}} \, \rho_0(k^*_{\sigma_0})\,e^{ 2\Gamma \,f(t)}.
\end{align}
Using Eq.~\eqref{fandg}, we find
\begin{align}
    n(t) \underset{t\rightarrow \infty}{\simeq} \frac{1}{\Gamma f(t)} \frac{\rho_0(k^*_{\sigma_0})^2}{4\pi n(0)} \simeq  \frac{1}{\Gamma g(t)} \frac{\rho_0(k^*_{\sigma_0})^2}{4\pi n(0)} ,
\end{align}
where, in the second identity, we have used the fact that the logarithmic term in Eq.~(\ref{fandg}) is subleading. Since $n(t) = \partial_t g(t) $, we arrive at an ordinary differential equation of the form $\partial_t g(t) \propto 1/g(t)$. Consequently, $g(t)\propto \sqrt{t}$, and then
\begin{equation}\label{longtime_2bdens}
    n(t) \propto t^{-1/2},
\end{equation}
as expected from our numerical results, see Fig.~\ref{density_2body}. We note that a similar result was found recently in a lattice gas  with similar but different two-body loss term~\cite{rossini_strong_2021} as well as in its continuous analog~\cite{rosso_one-dimensional_2022}, although we stress that the loss functionals and rate equations for these models are different from the ones of this paper. 

We conclude this subsection with an investigation of the long time behavior of the rapidity distribution $\rho(k,t)$, which is determined by the mean density $n(t)$ according to Eq.~\eqref{Eq:rhokt:2bl}. We have just we etablished that the first Fourier mode of the initial rapidity distribution strongly influences the long time behavior. In the case of a vanishing first Fourier mode, the rapidity density at time $t$ is simply given (see Eqs.~\eqref{Eq:tbl:mf}-\eqref{Eq:rhokt:2bl}) by $\rho(k,t) = \rho_0(k)\, n(t)/n(0)$. The ratio $\rho(k,t)/n(t)$ is then time-independent, as illustrated in Fig.~\ref{figlongtime2}(c). In contrast, when the first Fourier mode of the initial distribution $\rho_0(k)$ is non-zero, the ratio $\rho(k,t)/n(t)$ loses its dependence on the initial rapidity distribution at very long times. Indeed, in that case the rapidity distribution goes to a low-density, low-temperature, Boltzmann distribution of the form $\rho(k,t)/n(t) \simeq e^{\beta(t) \cos k} $ with the effective inverse temperature $\beta (t) = 2 \sigma_0 \Gamma f(t)$. This is illustrated for the case of an initial thermal rapidity distribution in Fig.~\ref{figlongtime2}(a), where we see that the ration $\rho(k,t)/n(t)$ gets  concentrated around $k=0$ and is very close to a Boltzmann distribution. 
Notice also that the effective temperature $\beta(t)$ is negative when the sign of the first Fourier mode of the initial rapidity distribution is negative. This is illustrated in Fig.~\ref{figlongtime2}(b), where we display the ratio $\rho(k,t)/n(t)$ at late time for the far-from-thermal initial rapidity distribution $\rho_0(k)=(1-\cos(k))/2$. We observe that, at late times, the distribution gets concentrated around $k=\pi$ and corresponds to a Boltzmann distribution at negative temperature.

Remarkably, these observations are in stark contrast with our findings for the $K=1$ case. While we found that, for $K=1$, the rapidity distribution never goes to a thermal distribution at late time, here for $K=2$ the distribution goes to a low-density, low-temperature (possibly negative), thermal distribution. This is always true, except in the special case where the first Fourier mode of the rapidity distribution vanishes; in that case the rapididity distribution is simply rescaled by a factor $n(t)/n(0)$ under lossy evolution.

\begin{figure}[htb!]
    \centering
    \includegraphics[scale=0.56]{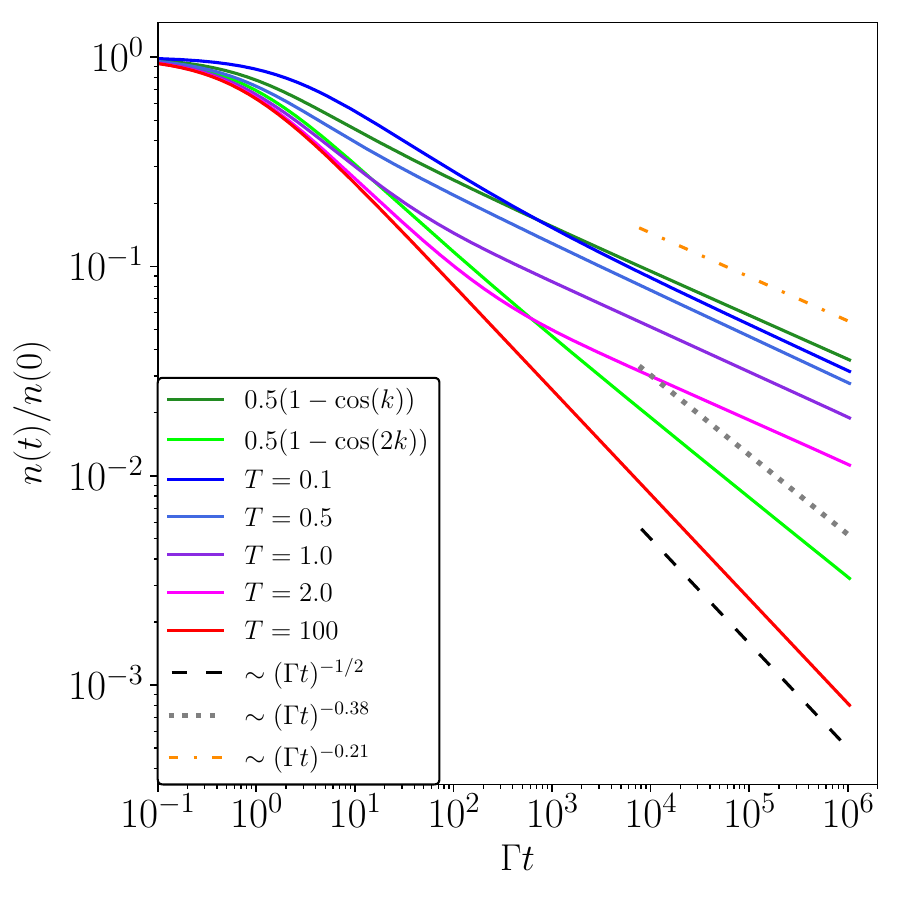}
    \caption{{\color{black} Time evolution of the boson density under three-body losses for different initial rapidity distributions at half filling. Colored curves are obtained by solving numerically the time evolution equation of $\rho(k)$ using the Runge-Kutta method with an non-regular time step. From blue to red, the simulation is performed with an initial distribution which is a Fermi-Dirac distribution $\rho
    (k,0)=(1+\exp(-\cos(k)/T))^{-1}$. The dark and light green curves are respectively obtained from initial rapidity distributions $\rho
    (k,0)= (1-\cos(k))/2$ and $\rho
    (k,0)= (1-\cos(2k))/2$. The three dashed lines are a guide to the eye showing power-law decay for three different exponents.}}
    \label{3b_dens}
\end{figure}

\begin{figure}[htb]
    \centering
    \includegraphics[scale=0.565]{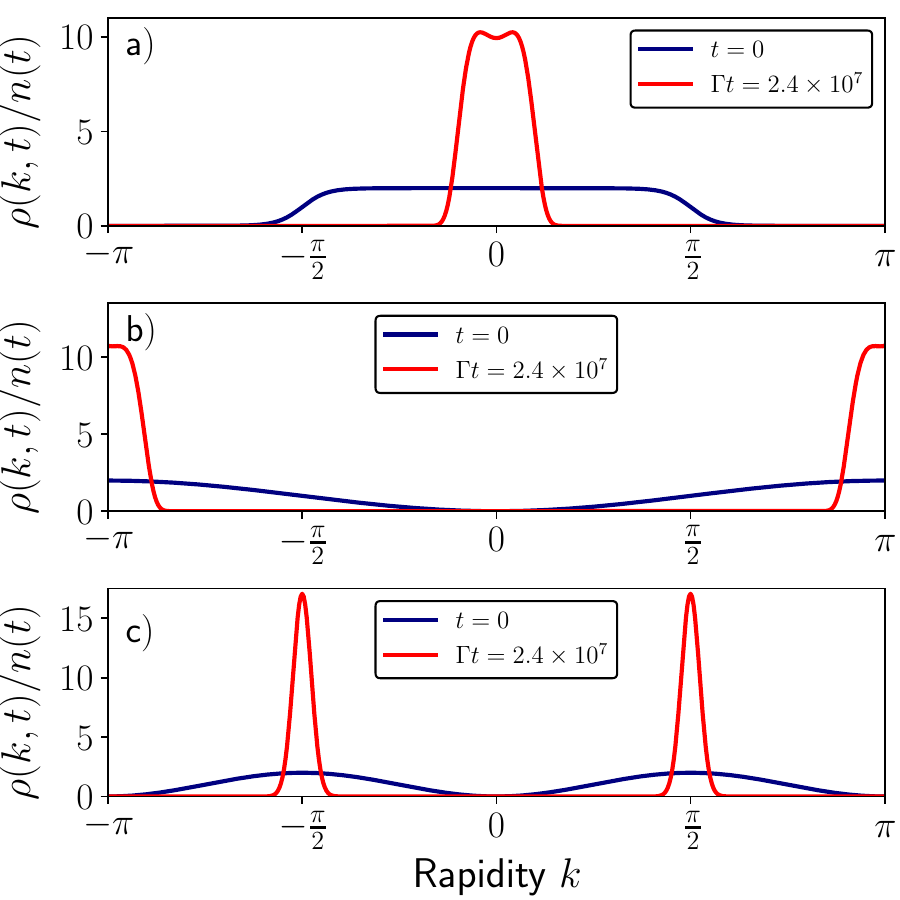}
    \caption{{\color{black}Long time behavior of different rapidity distributions rescaled by the corresponding density under three-body losses. For the subplots a, b and c, the blue curve represents the initial rapidity distribution while the red curve is the rapidity distribution at long time. a) The initial distribution is a Fermi-Dirac distributions $\rho_0
    (k)=1/(1+\exp(-\cos(k)/T))$ with $T$=0.1. b) Non-thermal initial rapidity distribution is $(1-\cos(k))/2$. c) Non-thermal initial rapididty distribution is $(1-\cos(2k))/2$.}}
    \label{figlongtime}
\end{figure}

\subsection{Generic observations for arbitrary $K$}

We now turn to the case of higher $K$, and draw some general conclusions.

Numerically, we solve the time evolution equation of the rapidity distribution for three-body losses ($K=3$), see Fig.~\ref{3b_kdis} for the evolution of the rapidity distribution from an initial thermal state, and Fig.~\ref{3b_dens} for the atom density $n(t)$. In Fig.~\ref{3b_kdis} we see that the effect of three-body losses is to spread the rapidity distribution in rapidity space, as already observed for one-body and two-body losses. We expect that this is a generic effect caused by $K$-body losses for any $K$. In Fig.~\ref{3b_dens}, we observe that the mean density decays as $t^{-1/2}$ for an initial infinite temperature state, while for any non-zero initial temperature it crosses over to a $t^{-\alpha}$ decay at long times with an exponent $\alpha \simeq 0.21$. This exponent seems to be independent of the initial temperature as long as it is non-zero, see Fig.~\ref{3b_dens}. However, for an initial rapidity distribution that is far from thermal, such as for instance $\rho(k,t=0) = (1-\cos k)/2$ or $(1-\cos (2k))/2$, we find that the density also decays as a power-law at late time, although with a different exponent $\alpha$: the exponent is close to $0.21$ for $\rho(k,t=0) = (1-\cos k)/2$, and close to $0.38$ for $\rho(k,t=0) = (1-\cos (2k))/2$.

We have not been able to analytically derive the observed generic power-law decay for $K=3$ or for higher $K$, beyond the special case of the initial infinite temperature state. The latter case is easily understood because, for an infinite temperature state the rapidity distribution is constant, $\rho(k)=n$, and the equation \eqref{rho_evo} can be solved analytically. Then the determinant of the matrix $B$ is equal to the determinant of $C$, and the matrices $A$ reduce to identical and diagonal matrices. Therefore the loss functional is simply given by
\begin{align}\label{Floss_flat}
F_{K}[\rho](k)=Kn^K,
\end{align}
which is the result expected from the mean-field approach. Then the solution of the evolution equation \eqref{rho_evo} gives the mean density
\begin{ceqn}
\begin{align}\label{n(k)K}
n(t)=\dfrac{n(0)}{(1+n(0)^{K-1} \, K(K-1) \, \Gamma t)^{1/(K-1)}}. 
\end{align}
\end{ceqn}
Beyond that simple case, we have not been able to express the loss functional in a simple form so as to derive the long-time decay of the mean density.

Similarly to the $K=1$ and $K=2$ cases, we have investigated the behavior of the rescaled rapidity distribution $\rho(k,t)/n(t)$ at late times. Recall that this ratio reveals that the gas generically (i.e. unless the first Fourier mode of $\rho(k)$ is tuned to zero) goes to a ,low-density, low-temperature thermal state for $K=2$, while for $K=1$ it never does. In Fig.~\ref{figlongtime} we display this ratio at late time for $K=3$. Fig.~\ref{figlongtime}.(a) corresponds to a thermal initial rapidity distribution, and Figs.~\ref{figlongtime}.(b)-(c) to non-thermal initial rapidity distributions $\rho(k,t=0)=(1-\cos k)/2$ and $(1-\cos (2k))/2$ respectively. We observe that the rescaled rapidity distribution concentrates around the maxima of the initial rapidity distribution at long times. Even for an initial thermal distribution (Fig.~\ref{figlongtime} (a)), the long time behavior of the density profile can not be described by a Boltzmann distribution, as it looks like a bell-shaped distribution that has a small dip at $k=0$. A similar conclusion holds for Fig.~\ref{figlongtime}.(b). Finally Fig.~\ref{figlongtime}.(c) shows the emergence of peaks localised at $k=\pm \pi/2$. We conclude that, in contrast with the $K=2$ case, the late-time rapidity distribution is generically non-thermal.


\section{Harmonically trapped gas}
\label{sec:trapped}

In many cold atom experiments, the gas lies in a longitudinal trapping potential. This prompts us to study the influence of the trapping potential on the dynamics of our lossy lattice hard-core gas. For simplicity we restrict to a harmonic potential $V(x)=\omega^2x^2/2$.

We adopt a coarse-grained perspective of the gas: we assume that the gas can be divided into fluid cells which contain a large number of bosons, and that the state of the gas within each fluid cell $[x, x+dx]$ is a certain macrostate represented by the local density of rapidities $\rho(x,k)$. Such coarse-grained descriptions have been very successful lately in describing the out-of-equilibrium quantum many-body dynamics of nearly integrable gases~\cite{bertini_transport_2016,castro-alvaredo_emergent_2016,doyon_lecture_2020,schemmer2019generalized,malvania2021generalized,bouchoule_generalized_2021,coppola_speculations_2023}. Here we investigate the effect of losses on our lattice hard-core gas within that coarse-grained description.

The equation satisfied by the position-dependent rapidity distribution is
\begin{align}\label{evo_Wig}
    &\partial_t \rho(x,k,t)+\sin(k) \, \partial_x \rho(x,k,t)\nonumber -\omega^2 x\, \partial_k \rho(x,k,t) \\ 
    &= -\Gamma F[\rho(x,.,t)](k).
\end{align}
In the first line, the term $\partial_x  \sin(k) \rho(x,k)$ corresponds to the gradient of the current of quasi-particles with rapidity $k$, $j(x,k) = \sin(k) \rho(x,k) $. Here $\sin(k)$ is the group velocity of quasi-particles with lattice dispersion relation $\varepsilon(k) = - \cos (k)$. The term $-\omega^2 x\, \partial_k \rho(x,k) $ in Eq.~(\ref{evo_Wig}) corresponds to Newton's second law, and encodes the fact that the quasi-particles feel the harmonic potential and are accelerated according to $\dot{k} = - \partial_x V(x) = - \omega^2 x$. Finally, the r.h.s of Eq.~(\ref{evo_Wig}) is the loss term at position $x$, which follows from the assumption that the gas is locally homogeneous so that we can apply the formalism developed in previous sections, this time within each fluid cell $[x,x+dx]$.

\begin{figure*}[ht]
    \centering
    \includegraphics[scale=0.585]{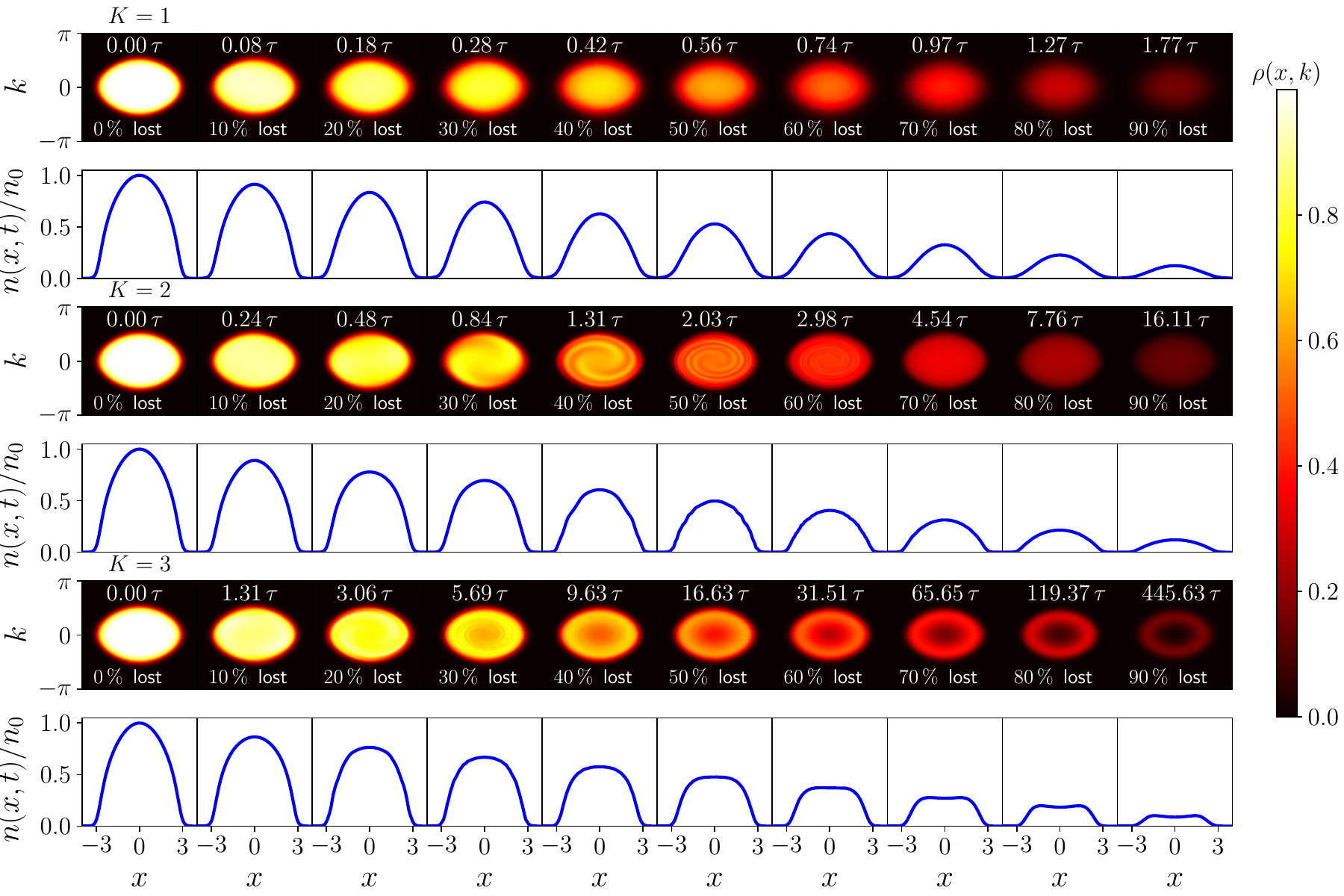}
    \caption{{\color{black} Evolution of the position-dependent distribution $\rho(x,k)$ in a harmonic trap and under $K$-body losses. The initial distribution is $\rho(x,k) = 1/(1+\exp((-\cos(k)+ \omega^2 x^2/2 )/T))$ with harmonic trap frequency $\omega$=1/2 and temperature $T$=0.1. The loss rate is $\Gamma$=0.1. The different snapshots show increasing times in units of $\tau = 2\pi / \omega$. Below each phase portrait we add the corresponding density profiles $n(x,t)$ obtained by integrating $\rho(x,k,t)$ over $k$. The density profiles are rescaled by the initial density at the trap center $n_0 = n(x=0,t=0)=0.5$. The two first rows show the effects of one-body losses ($K$=1) on the distribution in phase-space and the corresponding density profiles. The two next rows show the effects of two-body losses ($K$=2) and the last two rows show the effect of three-body losses ($K$=3). Numerically, the $x$ and $k$-axis are discretized with 300 points on each axis, and we use a time step of $\Delta t = 0.05$; we provide a check of convergence of the numerical method in Appendix~\ref{check_num}.}}
    \label{phase_space_loss}
\end{figure*}

\subsection{Numerical method}

Our main goal in this section is to solve numerically the evolution equation (\ref{evo_Wig}). For this we use a split-step method. Assuming that we know the rapidity distribution $\rho_t (x,k)$, from time $t$ to $t+\Delta t$ we first compute the new rapidity distribution $\rho'_{t+\Delta t} (x,k)$ generated by the transport of quasi-particles, and then compute $\rho_{t+\Delta t} (x,k)$ from $\rho'_{t+\Delta t} (x,k)$ by implementing localized lossy evolution during a time step $\Delta t$.

{\color{black} This gives the following scheme. For the transport part we rely on the fact that the underlying dynamics is the one of non-interacting quasi-particles. Each quasi-particle at position $(x,k)$ in phase space evolves as
\begin{equation}\label{xk_traj}
    \left\{ \begin{array}{rcl}
    \dfrac{dx}{dt}=\sin(k) \\ \\
    \dfrac{dk}{dt}=-\omega^2 x .
    \end{array} \right. 
\end{equation}
Let us call $X(x,k, t)$ and $K(x,k,t)$ the analytical solution of that equation. Then, 
the first step of our numerical scheme is
\begin{eqnarray}
\label{step1}
  \rho'_{t + \Delta t} ( X(x,k,t) ,K(x,k,t) ) &=& \rho_{t} (x,k) ,
\end{eqnarray}
and the second step is
\begin{equation}\label{step2}
    \rho_{t + \Delta t}(x,k) = y (\Delta t, k),
\end{equation}
where $y(\tau,k)$ is the solution of the differential equation
\begin{equation}\label{y_diff}
         \partial_\tau y(\tau,k) = - \Gamma F[ y(\tau,.) ](k) , \qquad y(0,k) = \rho'_{t+\Delta t, x} .
\end{equation}
The rapidity distribution $\rho (x,k)$ is discretized on a regular grid in phase space, and the two steps in Eq.~(\ref{step1}) and Eq.~\eqref{step2} are implemented as follows.

For the transport step in \eqref{step1}, we numerically integrate Eq.~(\ref{xk_traj}) for each point of the grid. Then, 
starting from the values $\rho_t(x,k)$ on the regular grid in phase space, we move each node of the grid according to $(x,k) \rightarrow ( X( x,k, \Delta t), K( x,k, \Delta t) )$. This gives us the rapidity distribution after transport over a time $\Delta t$, which however is defined on a new grid, different from the initial regular grid. To get the new rapidity distribution $\rho'_{t+\Delta t}(x,k)$ on the initial regular grid, we use linear interpolation. Notice that the whole procedure, including both the numerical integration of (\ref{xk_traj}) and the linear interpolation, needs to be performed only once. To benchmark this transport part, we have checked that this method gives excellent numerical precision for the simulated transport in the absence of losses.}

The second step consists in solving numerically the differential equation~\eqref{y_diff} for each column of the grid in phase-space. The solution of each differential equation is obtained by the Runge-Kutta method with the initial condition $y(0,k) = \rho'_{t+\Delta t, x}$. We do this for each column of the grid, and we thus get the new space-dependent rapidity distribution $\rho_{t+\Delta t}(x,k)$ according to Eq.~\eqref{step2}.

The combination of the two steps allows us to go from a rapidity distribution $\rho_{t}(x,k)$ to the new distribution $\rho_{t+\Delta t}(x,k)$, both defined on the same phase-space grid. We then repeat this procedure many times with a small time step $\Delta t$ to simulate the lossy evolution of the gas in the trap.

\subsection{Results}

We have performed numerical simulations of the evolution of the position-dependent rapidity distribution $\rho(x,k)$ under $K$-body losses using the algorithm presented in the previous section, see Fig.~\ref{phase_space_loss}. For the initial state, we use a thermal (Fermi-Dirac) rapidity distribution $\rho(x,k)=1/(1+\exp{(-\cos(k)+ \omega^2 x^2/2-\mu)/T})$, where $\mu$ is the chemical potential and $T$ the temperature.

Our numerical study allows us to make the following general observations, illustrated in Fig.~\ref{phase_space_loss}. For all $K$ (we have simulated $K=1,2,3$) the distribution typically spreads in phase-space, similarly to the homogeneous case. However, while the number of particles decays exponentially for $K=1$, the loss dynamics is much slower for higher $K$, and this has visible effects on the distribution of rapidities after a given percentage of lost atoms. 

For $K=1$ we observe that 
the edges of the phase-space distribution get depopulated very fast and quickly results in a halo around the origin. The spreading is also clearly visible in real space in the particle density profile $n(x) = \int \rho(x,k) dk/(2\pi)$, see the second line of Fig.~\ref{phase_space_loss}. 

For $K=2$, the situation is a little bit different, see Fig.~\ref{phase_space_loss}. Until $\sim$20$\%$ of the atoms have been lost, the dynamics is similar to the one for $K=1$, but after that we observe the formation of spirals in the bulk of the phase-space distribution. The spirals become more visible as the ratio $\Gamma/\omega$ is increased, see Appendix~\ref{app:spirals}. Compared to $K=1$, the bulk of the distribution also gets depopulated, leading to an approximately uniform circular droplet in phase-space, with a density that decays with time. This is visible in Fig.~\ref{phase_space_loss} after $\sim$50$\%$ of the atoms have been lost; we note that, compared to the case of one-body losses, the distribution spreads less significantly in phase-space.

\begin{figure}[th]
    \centering
    \includegraphics[scale=0.53]{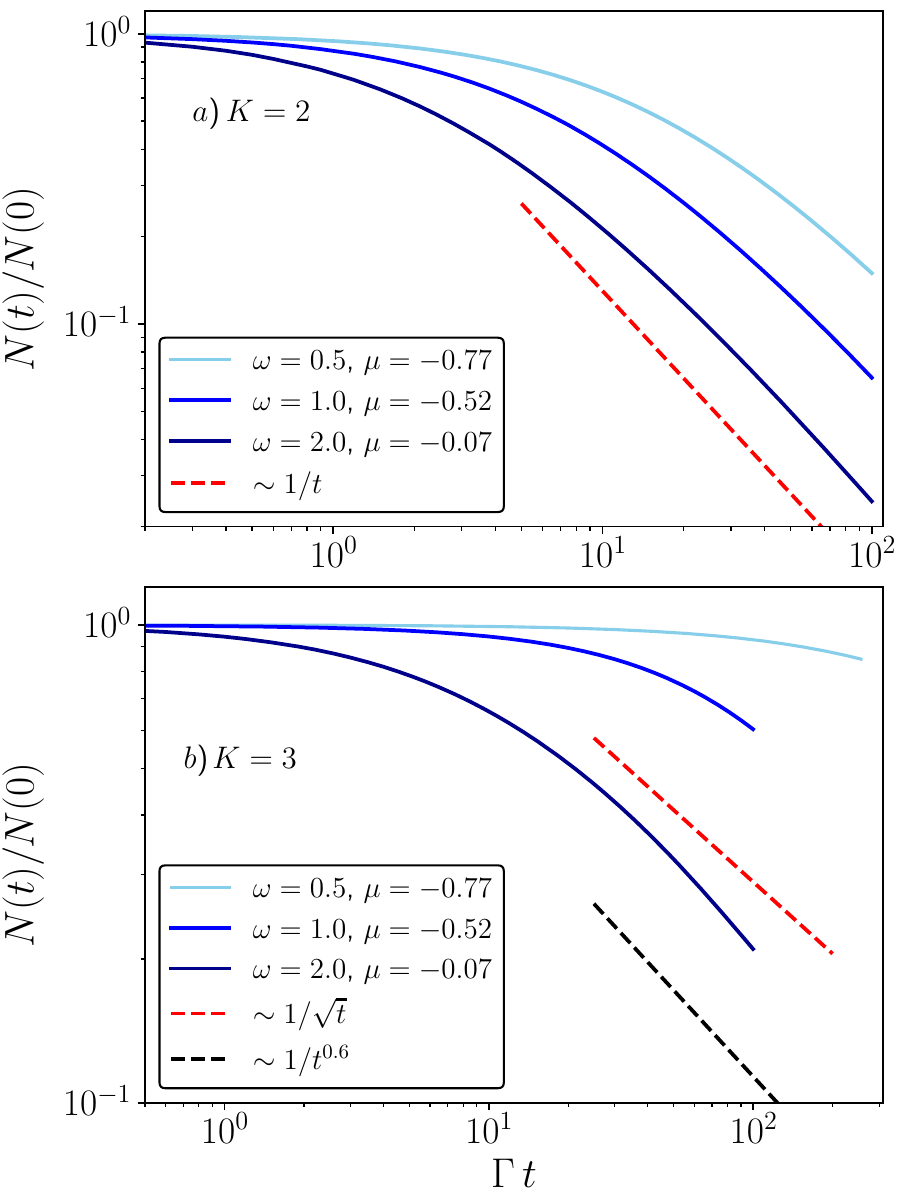}
    \caption{{\color{black} Time evolution of the total particle number under two-body losses. The initial distribution is $1/(1+\exp((-\cos(k)+ \omega^2 x^2/2-\mu)/T)$, where $T=0.1$ and $\mu$, the chemical potential, is chosen to fix the initial mean particle density to 0.5 and $N(0)$ represents the initial number of particle. The loss rate is $\Gamma$=0.1. (a) Results for two-body losses. The red dashed line is the long time behavior of the total particle number for the homogeneous case with an initial rapidity distribution which has no first Fourier mode (see Fig.(\ref{2bmany})). (b) Three-body losses. The black dashed line represents the asymptotic behavior of the mean density and the red dashed line is the mean-field prediction.}}
    \label{23bloss_pdens}
\end{figure}

For $K=3$, we also observe a small spiral appearing at the center of the distribution after $\sim$30$\%$ of the atoms have been lost. This spiral remains localised at the center during the dynamics. Interestingly, as one can see from the fifth line in Fig.~\ref{phase_space_loss}, this time it is the center of the phase-space distribution that decreases faster compared to the edges. After $\sim$60$\%$ of atoms lost, a hole starts developing at the center of the phase-space distribution, and the latter looks more and more like a ring. This is a clear signature of a strongly out-of-equilibrium gas in the trap, with a population inversion: the higher-energy single-particle orbitals get more populated than the low-energy ones. This effect is reflected in the corresponding real-space particle density $n(x)$ then acquires a doubly-peaked shape, with a local minimum at $x=0$, similarly to what we observed also in the homogeneous case.

Finally, having simulated the dynamics of the position-dependent rapidity distribution $\rho(x,k)$, one can easily get the evolution of the total particle number $N(t)$ by integrating $\rho(x,k,t)$ over $x$ and $k$ at fixed time $t$. For one-body losses, one always finds an exponential decay $N(t) = e^{-\Gamma t} N(0)$. For two-body and three-body losses the result is more interesting. In Fig.~\ref{23bloss_pdens}.(a), we show the evolution of the mean density under two-body losses for different trap frequencies $\omega$. Starting with a thermal distribution at temperature $T$ and chemical potential $\mu$, we observe that the mean density decreases at long times as $\sim 1/t$, a result that coincides with the mean-field (or infinite temperature) decay for the homogeneous gas. \textcolor{black}{Moreover, we see that the higher the trap frequency the faster the decay of the total particle number. Indeed, as the trap frequency is increasing, the distribution concentrates around the center of the confinement potential. Since the density of particles is higher, the probability of finding two neighboring bosons increases and the depopulation is then accelerated.}

For three-body losses, we see in Fig.~\ref{23bloss_pdens}.(b) that, like in the two-body case, a stronger confinement speeds up the decrease of total particle number. However, this time we do not observe a clear convergence towards the expectation from the mean-field (or infinite temperature) result for the homogeneous case, which would be $N(t) \sim t^{-1/2}$. We observe a number of particles that decreases approximately as a power-law $\sim t^{-\alpha}$ with an exponent $\alpha \simeq 0.6$. Our numerics does not allow us to draw a clear conclusion as to whether or not this would go to the mean-field exponent $1/2$ at longer times. \textcolor{black}{Nevertheless, let us stress that, qualitatively, the effect of three-body losses is the same as in the two-body case: compared to the homogeneous case, the trap dramatically speeds up the losses. This can be explained by the same mechanism as in the two-body case: since the density of particles is higher in the center, the probability of finding three neighboring bosons increases and the losses are enhanced.}

\section{Conclusion}
\label{sec:conclusion}
We have studied the effects of $K$-body losses on a gas of lattice hardcore bosons, in particular their effect on the thermalisation of the gas at late time. For this, we have relied on the hypothesis of adiabatic losses used previously in Refs.~\cite{bouchoule_effect_2020,lange_time-dependent_2018,Lange_2017,Lenarcic_2018, rossini_strong_2021}. We derived analytical results for the loss functional for any integer $K$ in the form of a small finite determinant, and closed expressions in the cases $K=1,2$.

For $K=1$ and $K=2$, we solved analytically the time evolution equation of the rapidity distribution of the spatially homogeneous gas. In the case of one-body losses, our formula~\eqref{result_rho} shows that the loss functional is in general non-linear and non-local in rapidity space, as already observed for the continuous Lieb-Liniger gas in Ref.~\cite{bouchoule_effect_2020}. After investigating the long time behavior of the rapidity distribution, we concluded that one-body losses do not drive the gas to a low-density thermal equilibrium state at long times. In the case of two-body losses, our formula~\eqref{Eq:rhokt:2bl} gives an implicit expression for the rapidity distribution and using a similar method as in Ref.~\cite{rossini_strong_2021}, we were able to investigate the long time behavior of the rapidity distribution and the mean particle density. In particular, we found that it decays generically as $\sim 1/\sqrt{t}$, except when the first Fourier mode of the initial distribution vanishes; in that case the particle density decays as $\sim 1/t$. A similar conclusion was drawn for a different loss process in Refs.\cite{rossini_strong_2021,rosso_one-dimensional_2022}.

Finally, we considered the inhomogeneous system consisting in a lattice hardcore bosons gas in harmonic potential. We provided a numerical method to solve the dynamics combining the effects of the losses and of the trapping potential. We observed that for $K\geq 2$ the trap generically speeds up the decay of the total particle number. We also found that the gas typically evolves towards a highly non-thermal state; in particular for $K=3$ we observe a striking ring-shaped distribution in phase space, which signals a inversion of population (i.e. higher energy single-particle orbitals get more populated than the lower energy ones), see Fig.~\ref{phase_space_loss}. Further investigations are needed to draw more quantitative conclusions about the dynamics in the trap.

\vspace{0.4cm}

{\bf Note:} while we were finishing this paper, a preprint by Perfetto, Carollo, Garrahan and Lesanovsky appeared \cite{perfetto2023quantum}, where a similar model of spinless fermions with $K=2,3,4$-body losses is studied within the context of quantum reaction-diffusion dynamics of annihilation processes. Our hard-core model coincides with their model for $K$ even, but not for $K$ odd because of the Jordan-Wigner mapping, as we explained in detail in Sec.~\ref{sec:lossfunctional}. Perfetto {\it et al.} do not focus on the effect of losses on the rapidity distributions of the gas, but rather on the evolution of the number of particles in the homogeneous setting.
For the case of $K$ even, where our models coincide, their findings about the evolution of the number of particles in the homogeneous setting are in agreement with ours.

\begin{acknowledgments}
    We thank Alberto Biella, Isabelle Bouchoule, Mario Collura and Leonardo Mazza for very useful discussions and for joint work on closely related topics. The work of JD, FR and DK is supported by the Agence Nationale de la Recherche through ANR-20-CE30-0017-01 project
‘QUADY’ and ANR-22-CE30-0004-01 project `UNIOPEN'. L.R. acknowledges hospitality from LPCT during the completion of this work.
\end{acknowledgments}

\appendix
\onecolumngrid
\section{Deriving solution \eqref{result_rho}}\label{appendix A}
For one-body losses, the time evolution of the rapidity distribution $\rho(t,k)$ is given by
\begin{ceqn}
\begin{align}\label{eqrho_app}
	\partial_t \rho = -\Gamma \left( \rho -(\rho^2-\mathcal{H}(\rho)^2-n^2(t))+2n(t) \mathcal{H}'(\rho)\right),
\end{align}
\end{ceqn}
where $\Gamma$ is the loss rate. We introduced the Hilbert transform $\mathcal{H}(f(x))=\frac{1}{2\pi}\int^\pi_{-\pi} dy \frac{f(y)}{\tan(\frac{x-y}{2})}$ with $f(x)$ a periodic function. The mean density $n(t)$ is known: $n(t)=n_0e^{-\Gamma t}$.

Here the rapidity distribution is a $2\pi$-periodic real-valued function. From the rapidity distribution, we can construct a complex-valued function whose the imaginary part is the Hilbert transform of the real part: $Q=\rho(k)+i\mathcal{H}(\rho(k))$. Such a function is called an analytic signal and can be analytically continued to the upper half-plane: the function $Q(z)=\frac{i}{2\pi}\int^\pi_{-\pi}\frac{dq \; \rho(q)}{\tan(\frac{z-q}{2})}$ is well-defined for $\Im(z)>0$ and $\Re(z)\in [-\pi,\pi]$ and reduces to $\rho(k)$ on the real axis. 

Taking the Hilbert transform of \eqref{eqrho_app}
\begin{ceqn}
\begin{align}\label{hilbeqrho_app}
	\partial_t \mathcal{H}(\rho) = -\Gamma \left( \mathcal{H}(\rho) -\mathcal{H}(\rho^2-\mathcal{H}(\rho)^2)+2n(t) \mathcal{H}(\mathcal{H}'(\rho))\right)
\end{align}
\end{ceqn}
and adding \eqref{eqrho_app} and i \eqref{hilbeqrho_app}, one has
\begin{ceqn}
\begin{align}
	\partial_\tau Q(\tau,z) = - (Q(\tau,z)-i2n\partial_z Q(\tau,z)-Q^2(\tau,z)+n^2(\tau) ).
\end{align}
\end{ceqn}
We used some properties of the Hilbert transform: i) $\mathcal{H}(\mathcal{H}(f))=-f$, ii) $\mathcal{H}'(f)=\mathcal{H}(f')$. Moreover, since $Q^2(z)$ is analytic for $\Im(z)>0$, the function $\rho^2-\mathcal{H}(\rho)^2+i2\rho\mathcal{H}(\rho)$ is an analytic signal if and only if $\mathcal{H}(\rho^2-\mathcal{H}(\rho)^2)=2\rho\mathcal{H}(\rho)$.\\
Introducing the function $Y(\tau,z)=Q(\tau,z+i2n(\tau))$, one gets
\begin{ceqn}
\begin{align}
	\partial_\tau Y(\tau,z) = Y^2(\tau,z)-Y(\tau,z)-n^2(\tau) 
\end{align}
\end{ceqn}
This equation can be solved if one assumes $Y(\tau,z)=\alpha(\tau,z)\;e^{-\tau}$. Indeed, thanks to this trick the above equation reduces to
\begin{ceqn}
\begin{align}
	\partial_\tau \alpha(\tau,z) = (\alpha^2(\tau,z)-n^2_0)\; e^{-\tau}
\end{align}
\end{ceqn}
Putting all terms depending on $\alpha$ in the left-hand side, one has
 \begin{ceqn}
\begin{align}
	\int \dfrac{d\alpha}{\alpha^2-n^2_0} =  -e^{-\tau}+C_1,
\end{align}
\end{ceqn}
which leads to
 \begin{ceqn}
\begin{align}
	\alpha(\tau,z) = n_0\tanh(n_0\,e^{-\tau}+C_2).
\end{align}
\end{ceqn}
The initial condition $Y(0,z)=Y_0$ sets the constant: $C_2=\tanh^{-1}(Y_0/n_0)-n_0$. Thus, one can write
\begin{ceqn}
\begin{align}
	Y(\tau,z) &= n(\tau)\tanh(n_0(e^{-\tau}-1)+\tanh^{-1}(Y_0/n_0))\nonumber \\
	&=n(\tau)\left(\dfrac{\tanh(n_0(e^{-\tau}-1))+Y_0/n_0}{1+\tanh(n_0(e^{-\tau}-1))Y_0/n_0} \right).
\end{align}
\end{ceqn}
Finally, the rapidity distribution reads
\begin{ceqn}
\begin{align}\label{result_tho_app}
	\rho(t,k)=n_0e^{-\Gamma t}\Re \left(\dfrac{\tanh(n_0(e^{-\Gamma t}-1))+\frac{i}{2\pi n_0} \int^\pi_{-\pi}dq \dfrac{\rho_0(q)}{\tan(\frac{k-q}{2}+in_0(1-e^{-\Gamma t}))}}{1+\frac{i}{2\pi n_0}\tanh(n_0(e^{-\Gamma t}-1)) \int^\pi_{-\pi}dq \dfrac{\rho_0(q)}{\tan(\frac{k-q}{2}+in_0(1-e^{-\Gamma t}))}}\right) .
\end{align}
\end{ceqn}

\section{Derivation of Eq.~\eqref{Eq:rhokt:2bl} of the main text}
\label{App:der:rhokt:tbl}
In this section we present the main steps to derive the exact expression of the rapidity distribution in the homogeneous case for $K=2$. We consider the time evolution of $n(t)$ written as:
\begin{equation}
    \partial_t n(t) = \dfrac{1}{2 \pi} \int_{-\pi}^{\pi} \partial_t \rho(k,t) dk.
\end{equation}
We now insert the evolution equation~\eqref{Eq:2bl:symm} obtaining:
\begin{align}
    \partial_t n(t)  = & - \dfrac{2 \Gamma n(t)}{2 \pi} \int_{-\pi}^{\pi} \rho(k,t) dk + \dfrac{\Gamma}{2 \pi^2} \left( \int_{-\pi}^{\pi} \cos{(k)} \rho(k,t) dk \right)^2 \nonumber \\
    = & - 2 \Gamma n(t)^2 + \dfrac{\Gamma}{2 \pi^2} \left( \int_{-\pi}^{\pi} \cos{(k)} \rho(k,t) dk \right)^2
\end{align}
By inverting the latter relation, one obtains:
\begin{equation}
   \left \lvert \int_{-\pi}^{\pi} \cos{(k)} \rho(k,t) dk \right \rvert = \pi \sqrt{\dfrac{2}{\Gamma} \left( \partial_t n(t) + 2 \Gamma n(t)^2 \right)},
\end{equation}
where $\lvert \bullet \rvert$ denotes the absolute value. At this point of the derivation it is useful to introduce the following variable: $\sigma (t) = \text{sgn} (\int_{-\pi}^{\pi} \cos{(k)} \rho(k,t) dk)$, with $\text{sgn} (x)$ being the sign function. We now claim that the function $\sigma (t) $ is solely determined by its value at initial time, i.e. $\sigma (t) = \sigma (0) \doteqdot \sigma_0$, the argument goes as follows. (i) If the first Fourier mode vanishes at some time $t$, then it must vanish also at any later time. This follows from Eq.~\eqref{Eq:2bl:symm}. (ii)  This implies that the sign of the first Fourier mode is continuous in time. Since it can take only discrete values, it is in fact a constant.

By inserting the latter equation in Eq.~\eqref{Eq:2bl:symm} the time evolution of the rapidity distribution can be then recasted into the following form:
\begin{equation}
  \dot \rho(k,t) = -2 \Gamma \, \rho(k) \, n(t) + \Gamma  \sigma_0
\cos{(k)} \rho(k) \sqrt{\dfrac{2}{\Gamma} \left( \partial_t n(t) + 2 \Gamma n(t)^2 \right)}.
\end{equation}
We now divide both sides by $\rho(k,t)$ and then integrate:
\begin{equation}
    \ln{\left(\dfrac{\rho(k,t)}{\rho_0(k)} \right)} = 
    -2 \Gamma \, \int_0^t n(t') dt' +  \Gamma  \sigma_0
\cos{(k)}  \int_0^t \sqrt{\dfrac{2}{\Gamma} \left( \partial_{t'} n(t') + 2 \Gamma n(t')^2 \right)} dt'.
\end{equation}
By exponentiating the latter equation we get:
\begin{equation}
    \rho(k,t) = \rho_0(k) \exp{- 2 \Gamma \int_0^t n(t') dt' +  \sigma_0 \cos{(k)} \int_0^t \sqrt{ \dfrac{2 \left( \partial_{t'} n(t') + 2 \Gamma n(t')^2 \right)}{\Gamma}} dt' } ,
\end{equation}
this concludes the derivation of Eq.~\eqref{Eq:rhokt:2bl}.

\section{Additional data for the homogeneous $K=2$ case}
\label{App:data:tbl}
\begin{figure}[H]
    \centering
    \includegraphics[scale=0.5]{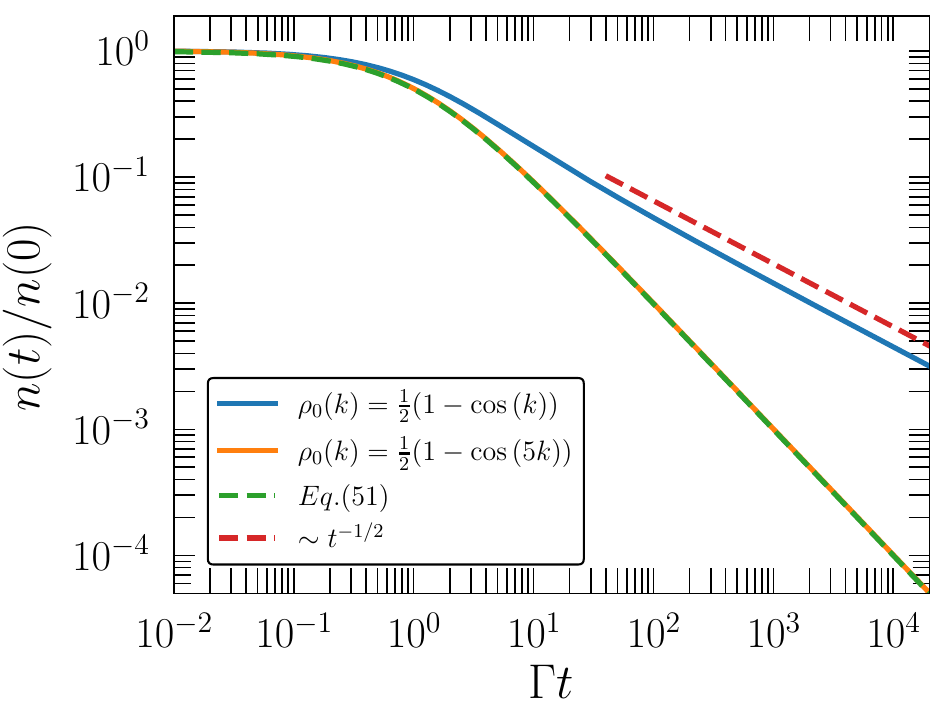}
    \includegraphics[scale=0.5]{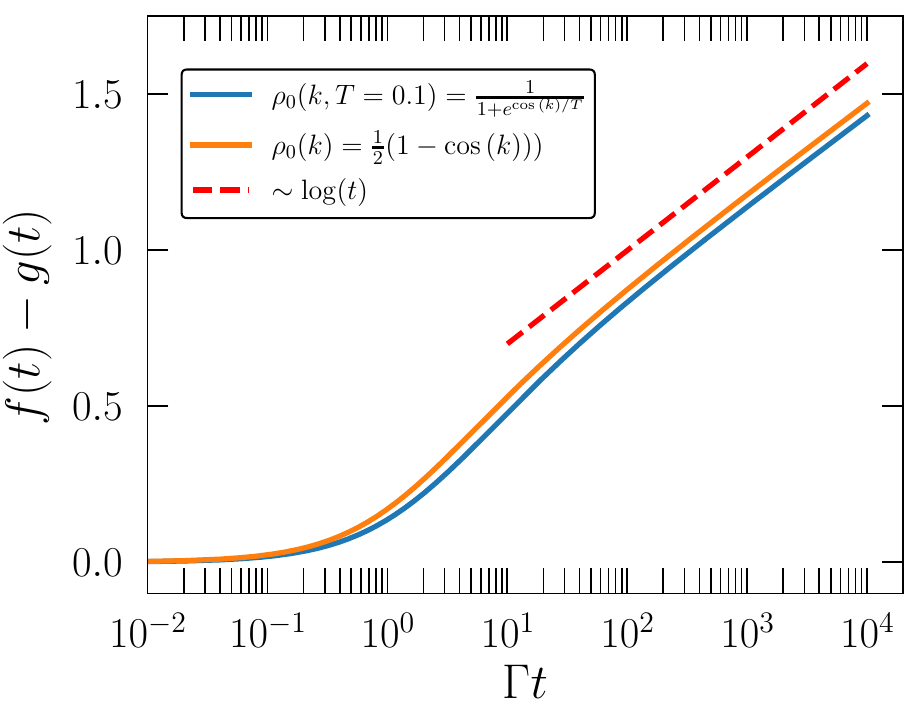}
    \caption{Left panel: the mean density under two-body losses for two different non-thermal rapidity distributions. Solid colored curves are obtained by solving numerically the time evolution equation of $\rho(k)$ for the loss functional~\eqref{2loss_F} with a time step $dt$=0.05 and a loss rate $\Gamma=$0.1. \textcolor{black}{The red dashed line is the long time limit of the mean density given by Eq.~\eqref{Eq:tbl:mf}. The long time behavior for non vanishing first Fourier mode distributions is presented by the black dashed line (see Eq.\eqref{longtime_2bdens})}. Right panel: difference between the two function $f(t)$ and $g(t)$ defined in the main text for a thermal and non-thermal distributions.}
    \label{Fig:tbl:data}
\end{figure}
In this section we present additional data concerning the homogeneous $K=2$ case. In particular, we show in Fig.~\ref{Fig:tbl:data} (left panel) the dynamics of the mean density for two different initial rapidity distributions which are not thermal. Firstly, we consider an initial distribution given by $\rho_0(k) = \dfrac{1}{2} (1 - \cos{(k)})$, which has first Fourier mode different from zero. Secondly, we consider $\rho_0(k) = \dfrac{1}{2} (1 - \cos{(5 k)})$, whose first Fourier mode vanishes. We see that the dynamics induced by the former distribution has a longtime behaviour guven by $\sim  1/ \sqrt{t}$, whereas the latter, due to its vanishining first Fourier mode, is described by Eq.~\eqref{Eq:tbl:mf}. This corroborates our findings for initial thermal distributions presented in the main text. Moreover, we show in Fig.~\ref{Fig:tbl:data} (right panel) the quantity $f(t) - g(t)$, where $g(t)=\int_0^t n(\tau ) \,d\tau$ and $f(t)=\int_0^t\sqrt{ 1 + \frac{\partial_{\tau} n(\tau)}{2 \Gamma \, n(\tau)^2}  }  n(\tau ) d\tau $ for two different rapidity distributions. In the main text we took the first order expansion of $f(t)$ resulting in a logarthmic growth for the quantity $f(t) - g(t)$, which is thus corroborated by the numerical data here presented. As such, given an initial rapidity distribution whose first Fourier mode is non-vanishing, one has a longtime decay of the mean density given by $n(t) \sim 1/ \sqrt{t}$.

\section{Brief discussion on the spiral in figure Fig.~\ref{phase_space_loss}}
\label{app:spirals}

During the evolution of the position-dependent rapidity distribution in phase space (see Fig.~\ref{phase_space_loss}) , the distribution exhibits a spiral, which is clearly visible after 40$\%$ of atoms lost for two-body losses. In principle, in a regime where the trap frequency $\omega$ is highly dominating the loss rate $\Gamma$, one expects that the distribution remains rotation invariant at all times. This implies that the spiral vanishes for $\omega \gg \Gamma$. To check this statement we compare the quantity $\rho(x,k)$ for two distincts values of $\omega$ (see the figure below). In figure Fig.~\ref{snapshot_app} We can see that the spiral appearing for $\omega=5\Gamma$ covers entirely the distribution, while for $\omega=20\Gamma$ the spiral is localised at the distribution's center. Moreover, in the case $\omega=20\Gamma$ we observe small oscillations between the edges and the center of the distribution. The frequency of these oscillations is high compared to the center of the distribution. 
\begin{figure}[H]
    \centering
    \includegraphics[scale=0.5]{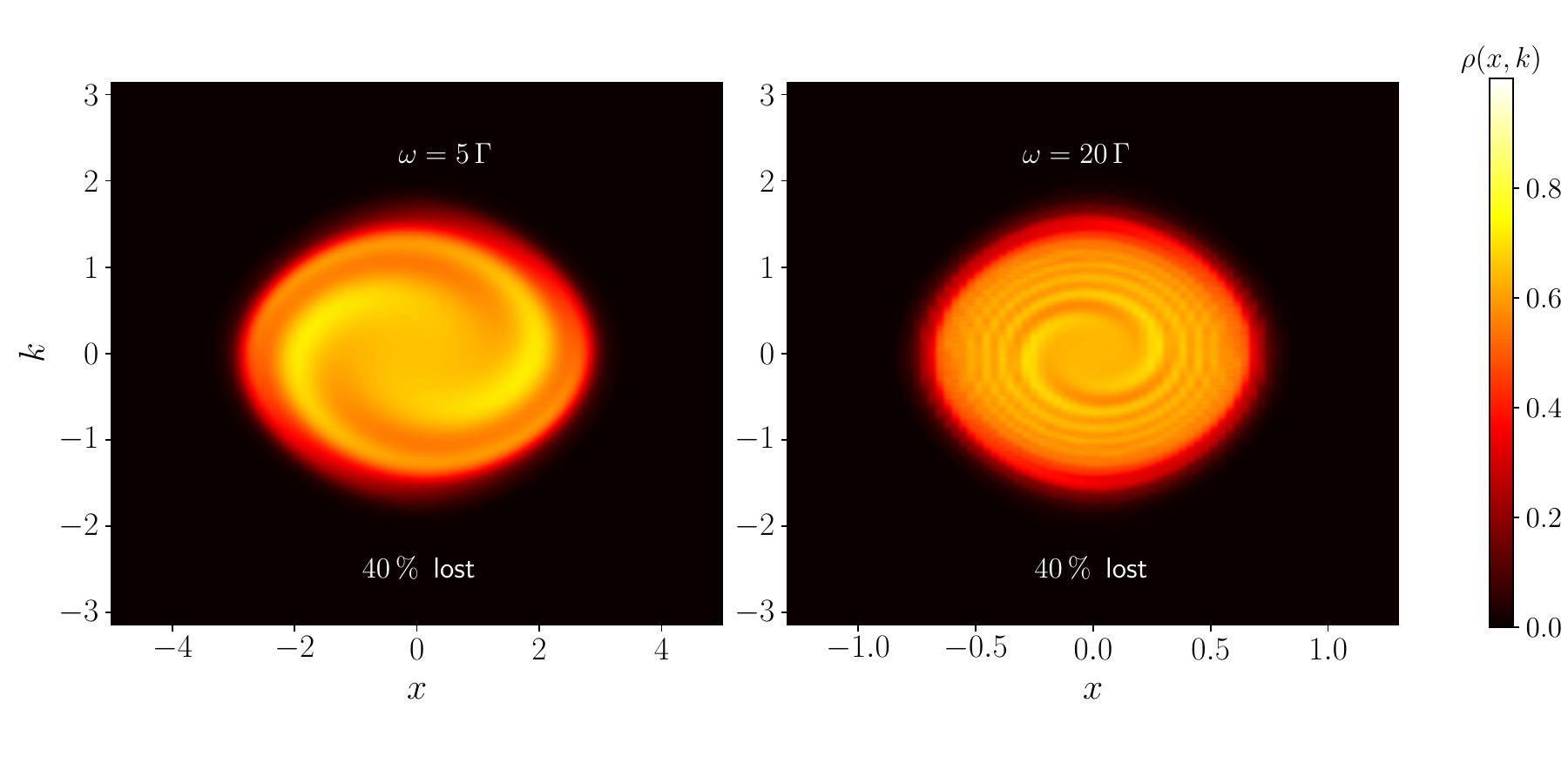}
    \caption{Phase portraits showing the spiral for two different values of $\omega$. The initial distribution is identical to the one used in figure Fig.~\ref{phase_space_loss} and here we only consider the distribution after 40$\%$ of atoms lost.}
    \label{snapshot_app}
\end{figure}

\section{Convergence of numerical results}

\label{check_num}
\begin{figure}[htb]
    \centering
    \includegraphics[scale=0.55]{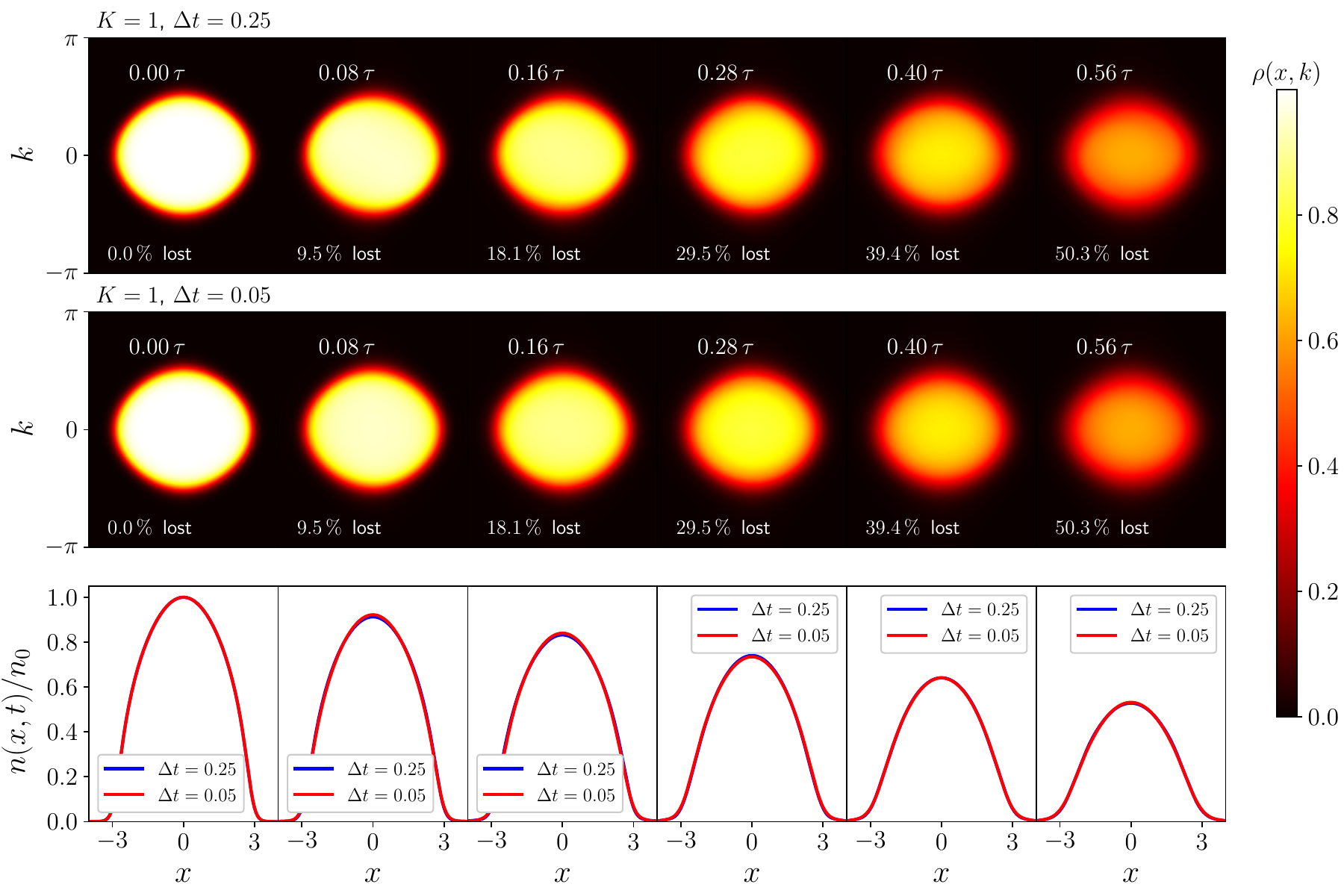}
    \caption{{\color{black} Dynamics of the local rapidity distribution $\rho(x,k)$ in phase space under one-body losses. The two first lines show the evolution of the distribution in phase space for two different time steps $\Delta t=0.25$ and $\Delta t=0.05$. The trap frequency is fixed to $\omega=1/2$, $\Gamma=0.1$ is the loss rate and $\tau$ is defined as $\tau=2\pi/\omega$. The last line corresponds to the density profils obtained by integrating over $k$ the local rapidity distribution. The numerical results are converged.}}
    \label{snapshot_appendix_k1}
\end{figure}

{\color{black} In this Appendix we provide evidence for numerical convergence of the results displayed in Fig.~\ref{phase_space_loss}. We have performed the same simulation as the one in Fig.~\ref{phase_space_loss} with a bigger numerical time step $\Delta = 0.25$ (instead of $\Delta = 0.05$). The physical parameters we use are (as in Fig.~\ref{phase_space_loss}):
\begin{align}
    \rho(x,k,t=0)=\frac{1}{1+\exp{(-\cos(k)+\omega^2x^2/2)T}},
\end{align}
with $T=0.1$ and $\omega=1/2$. We show results for one-body losses ($K=1$), with a loss rate $\Gamma$=0.1.

In Fig.~\ref{snapshot_appendix_k1}, we compare the results obtained for two different time steps $\Delta t=0.25$ (first row) and $\Delta t=0.05$. We see that the results are converged.
}

~\\
~\\
~\\
~\\
~\\
~\\
~\\
~\\
~\\
~\\
~\\
~\\
~\\
~\\
~\\
~\\
~\\
~\\
~\\
\twocolumngrid
\bibliographystyle{ieeetr}
\bibliography{Atom_losses_v2}
\end{document}